\documentstyle[10pt,aaspptwo,epsf]{article}
\lefthead{W. L. Holzapfel \ea }
\righthead{Ho A2163}
\def\t{{\tau}}
\def\b{{\beta}}
\def\D{{\Delta}}
\def\L{{\Lambda}}
\def\ea{{\it et al.}\ }
\def\sz{Sunyaev \& Zel'dovich}
\def\comp{Comptonization}
\def\pr{^{\prime}}
\def\2pr{^{\prime\prime}}
\def\ah{^{\rm h}}
\def\am{^{\rm m}}
\def\greatsim{\mathrel{\raise.3ex\hbox{$>$\kern-.75em\lower1ex\hbox{$\sim$}}}}
\def\lesssim{\mathrel{\raise.3ex\hbox{$<$\kern-.75em\lower1ex\hbox{$\sim$}}}}
\def\as{^{\rm s}}

\begin{document}

\title{Measurement of the Hubble Constant from X-ray and $2.1\,$mm
Observations of Abell 2163.}

\author{
W.~L.~Holzapfel\altaffilmark{1},
M.~Arnaud\altaffilmark{2},
P.~A.~R.~Ade\altaffilmark{3},
S.~E.~Church\altaffilmark{4},\\
M.~L.~Fischer\altaffilmark{5},
P.~D.~Mauskopf\altaffilmark{4,6},
Y.~Rephaeli\altaffilmark{7},
T.~M.~Wilbanks\altaffilmark{8}
and A.~E.~Lange\altaffilmark{4}}


\altaffiltext{1}{Enrico Fermi Institute, University of Chicago,
Chicago IL 60637}
\altaffiltext{2}{C. E. A., DSM, DAPNIA, Service d'Astrophysique, CE Saclay, 
F-91191 Gif sur Yvette Cedex, France}
\altaffiltext{3}{Department of Physics, Queen Mary and Westfield
College, Mile End Road, London, E1~4NS, U.K.}
\altaffiltext{4}{Department of Physics, Math, and Astronomy, California
Institute of Technology, Pasadena CA 91125}
\altaffiltext{5}{Energy \& Environment Division, Lawrence
Berkeley Laboratory, Berkeley, CA 94720}
\altaffiltext{6}{Department of Physics, University of California,
Berkeley, CA 94720}
\altaffiltext{7}{Center for Particle Astrophysics, University of California,
Berkeley, CA 94720}
\altaffiltext{8}{Aradigm Corporation, Hayward CA 94545}

\begin{abstract}
We report $2.1\,$mm observations of the Sunyaev-Zel'dovich (S-Z)
effect; these observations confirm our previous
detection of a decrement in the Cosmic Microwave Background intensity 
towards the cluster Abell 2163.
The S-Z data are analyzed using the relativistically correct 
expression for the \comp.
We begin by assuming the intracluster (IC) gas to be isothermal 
at the emission weighted average temperature determined by a
combined analysis of the ASCA and GINGA X-ray satellite observations.
The results of ROSAT/PSPC observations are used to determine an isothermal 
model for the S-Z surface brightness.
Fitting to this model, we determine the peak \comp\ to be 
$y_0=3.73^{+.47}_{-.61}\times 10^{-4}$.
The uncertainty includes contributions due to statistical uncertainty 
in the detection, instrumental baseline, calibration, 
and density model.
Combining the X-ray and S-Z measurements, we determine
the Hubble constant to be 
$H_0(q_0=\frac{1}{2})= 60^{+40}_{-23}\,{\rm kms}^{-1}{\rm Mpc}^{-1}$,
where the uncertainty is dominated by the systematic difference
in the ASCA and GINGA determined IC gas temperatures.
ASCA observations suggest the presence of a significant thermal
gradient in the IC gas.
We determine $H_0$ as a function of the
assumed IC gas thermal structure. 
Using the ASCA determined thermal structure
and keeping the emission weighted average temperature the same
as in the isothermal case, we find
$H_0(q_0=\frac{1}{2})= 78^{+54}_{-28}\,{\rm kms}^{-1}{\rm Mpc}^{-1}$.
Including additional uncertainties due to 
cluster asphericity, peculiar velocity, IC gas clumping, 
and astrophysical confusion, we find
$H_0(q_0=\frac{1}{2})= 78^{+60}_{-40}\,{\rm kms}^{-1}{\rm Mpc}^{-1}$.
\end{abstract}
%

\section{Introduction}
\label{cintro}
Compton scattering of the cosmic microwave background (CMB) radiation
by hot intracluster (IC) gas -- the Sunyaev-Zel'dovich (S-Z) effect --
gives rise to an observable distortion of the CMB spectrum which
can be used as a sensitive cosmological probe.
The intensity change caused by the scattering has a thermal 
component, due to the random motions of the scattering electrons
(\cite{SZ72}), and a kinematic component, due to the bulk
peculiar velocity of the IC gas (\cite{SZ80}).
When combined with X-ray measurements, the amplitude of the thermal
component can be used to determine the Hubble constant
(\cite{Gunn}; \cite{Cal79}).
The ratio of the amplitudes of thermal and kinematic components can be
combined with the X-ray temperature to determine the radial component
of the cluster peculiar velocity (\cite{SZ80}; \cite{RL}).
Because the surface brightness of the S-Z effect is independent of
redshift, this method for determining the Hubble constant and
cluster peculiar velocities can be applied to distant clusters, 
provided the requisite X-ray data can be obtained.

The thermal and kinematic components of the S-Z effect have distinct
spectral shapes and can be separated from one another by
measurements at two or more millimeter (mm) wavelengths.
Astrophysical confusion for the S-Z effect exhibits a broad minimum
at wavelengths of $2-3\,$mm (\cite{FL}).
Several significant detections of the S-Z effect at centimeter (cm) wavelengths
have been reported in recent years. Some of these measurements have been
combined with X-ray observations to compute values for $H_0$
(\cite{BHA}; \cite{BH94}; \cite{Herbig}; \cite{Jones}; \cite{Myers};
for a recent review, see \cite{R95a}).
The prospect of separating the thermal and kinematic
components of the S-Z effect while minimizing astrophysical
confusion makes observations at mm wavelengths attractive.

We have developed a novel instrument and observing strategy with
high sensitivity to the S-Z effect in several millimeter bands
(\cite{Holzapfel96a}).
Using this instrument, we have detected the S-Z effect in the direction
of the cluster Abell 2163 (\cite{W94} -- hereafter W94).
A2163 is a rich, moderately distant ($z= .201$, \cite{Soucail})
cluster of galaxies, which has been observed in the X-ray with
the Einstein, GINGA, ROSAT, and ASCA satellites (\cite{Arnaud92};
\cite{Elbaz}; \cite{Markevitch96}).
Its X-ray luminosity and temperature are among the highest of 
any cluster yet studied, implying a high IC gas
density and temperature, and correspondingly high S-Z surface brightness.
The low declination and modest angular extent of this cluster make it a
good candidate for observations with our instrument, which operates
by drift scanning (see Sections~\ref{Instrument} \& \ref{Observations}).
In this paper, we combine observations of the S-Z effect
and X-ray emission to determine $H_0$.
This is the first determination of $H_{0}$ from millimeter wavelength
measurements of the S-Z effect.

The determination of $H_0$ presented here includes
several refinements with respect to previous work.

1.) We use a relativistically correct calculation of CMB \comp\ in
clusters.
The nonrelativistic treatment is inaccurate at high
frequencies and high IC gas temperatures (see Section~\ref{TRcomp}).
Use of the relativistically correct formulae in the analysis 
of $2.1\,$mm measurements of A2163, the hottest known X-ray
cluster, is essential in order to derive an accurate value of $H_0$.

2.) We use the combination of the ROSAT derived density profile
and the S-Z surface brightness to constrain models for the large scale
thermal structure of the IC gas.
The allowed range of thermal structure models is compared to that 
determined from the combination of ASCA and ROSAT measurements.
We have determined $y_0$ and $H_0$ as a function of the 
cluster thermal structure, specifically for the 
thermal structure measured by ASCA.

3.) Finally, we include a detailed analysis of additional
contributions to the uncertainty in $H_0$ due to uncertainties
in the distribution and thermal structure of the IC gas, the
possibility that the cluster has an appreciable peculiar velocity,
and astrophysical confusion to the S-Z effect.

In Section~\ref{Theory}, we briefly review the relativistic \comp\
calculation and outline the method used to determine $H_0$ from
S-Z and X-ray measurements.
The analysis of the X-ray data is outlined in Section~\ref{Xanal}.
The SuZIE instrument, its calibration, and observations of A2163
are described in \ref{mmobs}.
In Section~\ref{2.1anal}, we describe the analysis of the S-Z data
under the assumption of isothermal gas.
We combine the isothermal S-Z and X-ray results to determine the 
Hubble constant in Section~\ref{hisogas}.
In Section~\ref{HTS}, we discuss the large scale thermal
structure of the IC gas and its effects on our results.
Additional contributions to the uncertainty in $H_0$ are discussed
in Section~\ref{Uncertainties}; our conclusions are summarized
in Section~\ref{Conclusion}.

\section{Theory}
\label{Theory}
\subsection{Relativistic \comp\ in Clusters}
\label{TRcomp}
The \sz\ (1972) treatment of CMB \comp\ by hot
IC gas is based on a solution to the
Kompaneets (1957) equation, a
nonrelativistic diffusion approximation to the full
kinematic equation for the change of the photon distribution.
The expression derived for the intensity change
-- an expression which has been used in virtually all works on the S-Z
effect in clusters -- is valid only at low gas temperatures.
A more accurate calculation of the thermal effect requires use of
the relativistic form for the electron
velocity distribution and the full expression for the scattering
probability, as has been presented by
Rephaeli (1995b).
Use of the correct relativistic calculation in this work
is essential because of the high
observing frequency and the unusually high IC gas temperature in A2163.

In the nonrelativistic limit, the CMB intensity change due to Compton
scattering by a hot electron gas (with no bulk velocity) is (\cite{ZS68})
\begin{equation}
\D I_{nr} = I_{0}\, y\, g(x)\, ,
\end{equation}
where $x=h\nu/kT_{0}$, $T_{0}$ is the CMB temperature, and
\begin{equation}
I_{0} \equiv \frac{2(kT_{0})^3}{(hc)^2}\,.
\end{equation}
The spectrum is given by the function,
\begin{equation}
g(x) = {x^4e^x \over \left(e^x -1\right)^2} \left[{x \left(e^x +1\right)\over e^x -1} -4\right] ,
\end{equation}
which vanishes at $x_0=3.83$ ($\nu_0=217\,$GHz) for $T_0=2.726\,$K (\cite{Mather}).
The \comp\ parameter is given by,
\begin{equation}
y = \int \left(\frac{kT_e}{mc^2}\right) n_e \sigma _T dl\,,
\label{y}
\end{equation}
where $n_e$ and $T_e$ are the electron density and temperature,
$\sigma _T$ is the Thomson cross section, and the integral is over a line
of sight through the cluster.

\begin{figure}[htb]
\plotone{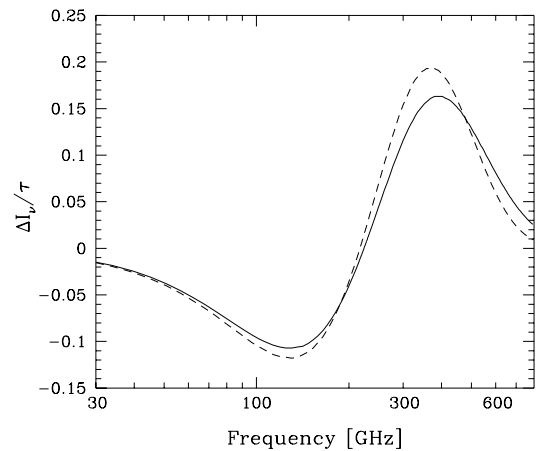}
\caption[]
{The spectral forms of the intensity change due to the
thermal Sunyaev-Zel'dovich effect and the non-relativistic
approximation. The solid line shows $ \D I/\t$,
and the dotted line shows $\D I_{nr}/\t$ (both in units of
$2(kT_{0})^3/(hc)^2$) for $kT_e = 14.6\,$keV.}
\label{Rcomp}
\end{figure}

As has been shown by
Wright (1979), the exact scattered photon 
distribution can be
calculated using the Chandrasekhar (1950)
expression for the scattering probability and the
relativistically correct form for the
Maxwellian electron velocity distribution. Integrations over the
scattering angle, electron velocities, and incident (Planck)
spectrum, yield an expression for the intensity
change (\cite{R95b}),
\begin{equation}
\D I =  I_{0}\int \Psi (x, T_e)\, d\t \, ,
\label{ISZ}
\end{equation}
where $d\t = n_e \sigma _T dl$.
The spectrum is given by
\begin{equation}
\Psi(x,\, T_e)=\frac{x^3}{\left(e^x-1\right)}\left[\Phi\left(x,T_e\right)-1\right] \,, 
\end{equation}
where $\Phi(x\,,T_e)$ is a three dimensional integral
over electron scattering angles, velocities, and the change
in frequency of the scattered photons which is
fully specified in Rephaeli (1995b).
This calculation
is exact to first order in $\t$. 
In Fig.~\ref{Rcomp}, we compare the relativistic and nonrelativistic expressions,
at the GINGA + ROSAT derived isothermal gas temperature
in A2163, $kT_e = 14.6\,$keV (\cite{Elbaz}).

At the central frequency of the SuZIE $2.1\,$mm band ($\nu \simeq 142\,$GHz),
the ratio of the brightnesses is $\D I/\D I_{nr}= .91$.
Finally, in order to facilitate comparison of our results with low
frequency observations, we combine equations~\ref{y} and \ref{ISZ}
to express an effective \comp\ in terms of the measured intensity change,
\begin{equation}
y =  \frac{\D I}{I_0}\, \frac{\int{ \frac{kT_e}{m_ec^2} d\t}}
{\int{ \Psi (x, T_e) d\t}} \, .
\label{yeff}
\end{equation}
All results for the \comp\ presented in this paper are
computed using this expression.

\subsection{Hubble Constant from S-Z and X-Ray Measurements}
The determination of $H_0$ from the combination
of X-ray and S-Z data has been treated by
Birkinshaw, Hughes, \& Arnaud (1991) -- hereafter 
BHA.  
Their treatment is
adopted here with some exceptions; we express the S-Z
decrement in intensity rather than in Rayleigh-Jeans temperature
and use a relativistically correct expression for the \comp.

Hot IC gas is detected primarily through its thermal bremsstrahlung
X-ray emission. The X-ray surface brightness at energy $E$ can be 
expressed as the line of sight integral (BHA equation 3.1),
\begin{equation}
b_X = \frac{1}{4 \pi \left(1+z\right)^3} \int {n_e^2 \L(E,\,T_e) dl}\,,
\label{BX}
\end{equation}
where $n_e$ is the electron density, $\L(E,\,T_e)$ is the specific
spectral emissivity, and $z$ is the cluster redshift.
We express the electron temperature, density, and
spectral emissivity in terms of the product of a reference value
and a dimensionless form factor (BHA equations 3.3 - 3.5):
\begin{equation}
n_e(r) = n_{e0}f_n(\theta,\,\phi,\,\xi)\,,
\end{equation}
\begin{equation}
T_e(r) = T_{e0}f_T(\theta,\,\phi,\,\xi)\,,
\end{equation}
\begin{equation}
\L_e(E,\,T_e) = \L_{e0}f_{\L}(E,\,T_e)\,,
\end{equation}
where $\theta$ is the angle from a reference line of sight, $\phi$ is the
azimuthal angle about that line of sight, $\xi$ is the angular
distance along the line of sight, and $n_{e0}$ and $T_{e0}$ are the
central gas density and temperature. We extend this treatment to
include a term describing the temperature dependence of relativistic
\comp,
\begin{equation}
\Psi(x,\,T_e) = \Psi_0 f_{\Psi}(x,\,T_e)\,.
\end{equation}
The expressions for the X-ray and S-Z surface brightnesses,
equations~\ref{BX} and \ref{ISZ}, then become:
\begin{equation}
b_X(\theta,\phi) =  \frac{ n^2_{e0} \L_{e0} d_A }
{4 \pi (1+z)^3 }\int{ d\xi f_n^2 f_{\L}}\,,
\label{X-surf}
\end{equation}
\begin{equation}
\equiv N_X \Theta_X\,,
\end{equation}
\begin{equation}
\D I \left(\theta,\phi\right) = I_0 \Psi_0 n_{e0}
\sigma_T d_A \int{d \xi f_n f_{\Psi}}\,,
\label{SZ-surf}
\end{equation}
\begin{equation}
\equiv N_{SZ} \Theta_{SZ} \,,
\label{SZ-surf1}
\end{equation}
where $d_A$ is the angular diameter distance.
The structural information for the cluster is contained in the angles:
\begin{equation}
\Theta_X = \int{d\xi f_n^2 f_{\L}}\,,
\end{equation}
\begin{equation}
\Theta_{SZ} = \int{d\xi f_n f_{\Psi}}\,.
\end{equation}
The normalization factors are then defined as:
\begin{equation}
N_X = \frac{n_{e0}^2 \L_{e0} d_A}{4 \pi \left(1+z\right)^3} \,,
\end{equation}
\begin{equation}
N_{SZ} = I_0 \Psi_0 n_{e0} \sigma_T d_A  \,.
\end{equation}
Eliminating the unknown central electron density between these
equations, we can express $d_A$ in terms of measurable quantities,
\begin{equation}
d_A = \left(\frac{N_{SZ}^2}{N_X}\right) \frac{\L_{e0}}{4 \pi
\left(1+z\right)^3  \left[ I_0 \Psi_0 \sigma_T \right]^2}\,,
\label{ANG DIA}
\end{equation}
which is an implicitly expressed generalization of BHA's equation
(3.12) valid at all frequencies and gas temperatures.
Combining $N_X$ and $N_{SZ}$ with a measurement of $T_{e0}$,
which appears in the expressions for $\L_{e0}$ and $\Psi_0$, we
can then solve for $H_0(q_0) \propto 1/d_A$.

Because we have no information about the distribution of the IC gas
along the line of sight,
we are required to make some simplifying assumptions about the
temperature and density form factors.
We assume that the density form factor can be described by a
convenient empirical model (\cite{Cal76}),
\begin{equation}
f_n(\theta) = \left(1+\frac{\theta^2}{\theta_c^2}\right)^{- \frac{3}{2} \b} ,
\label{neq}
\end{equation}
in which $\b$ and the cluster angular core radius ($\theta_c$) 
are left as free parameters.
The thermal structure of the IC gas is also assumed to be radially 
symmetric.
Together these assumptions make possible the determination of $f_n$
and $f_T$ from the observed X-ray
surface brightness as a function of X-ray energy.

Because $\L(E,\,T_e)$ and $\Psi(x,\,T_e)$ are non-linear functions of
the temperature, the shapes of the X-ray and S-Z surface brightness
profiles depend not only on the temperature form factor but also on its
reference value.
In general, the form factors cannot be determined from monochromatic
imaging data if the cluster gas is not isothermal.
However, at energies much lower than the
electron thermal energy (as in the ROSAT/PSPC energy band), the
X-ray emissivity of the IC gas depends weakly on 
the IC gas temperature.
Therefore, even for a non-isothermal IC gas, a good approximation 
to the density form factor can be
determined from the X-ray surface brightness profile.

The S-Z surface brightness is roughly proportional to the
electron pressure integrated along the line of sight through the cluster.
It is possible to constrain spherically symmetric models for $f_T$
by integrating the product of $f_n$ (determined consistently from
the X-ray data for each  $f_T$ model)
and $f_\Psi(x,T_e)$ along the line of sight to determine a model for the S-Z
surface brightness, which is then fit to the $2.1\,$mm data.
In  Section~\ref{HTS}, we use this method to constrain
models for the IC gas thermal structure.
Presently, much stronger constraints on the thermal structure are 
determined from the spatially resolved X-ray spectra
obtained by the ASCA satellite.

\section{X-ray Analysis}
\label{Xanal}

A2163 was the target of pointed observations by the GINGA and ROSAT
satellites. 
A detailed combined analysis of these observations has been published
previously (\cite{Elbaz}).
The ROSAT/PSPC and GINGA spectra were simultaneously
fit to a redshifted isothermal plasma emission model where the hydrogen
column density along the line of sight ($N_{\rm H}$), heavy metal abundance,
total emission measure, and IC gas temperature were free parameters.
The best fit emission weighted temperature was found to be 
$T_e =14.6^{+.9}_{-.8}\,$keV,  at $90\%$ confidence (\cite{Elbaz}).

\begin{figure}[htb]
\plotone{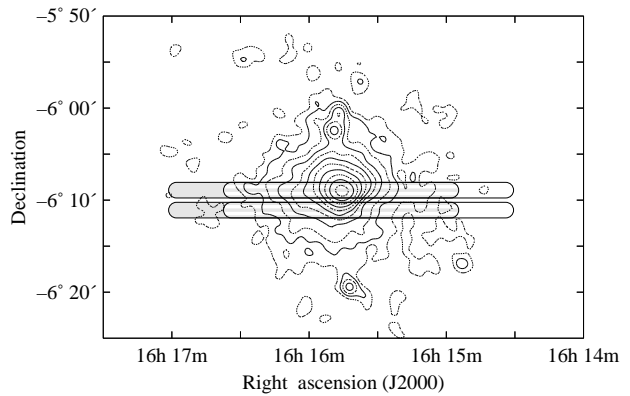}
\caption[]
{X-ray image of the A2163 cluster observed
with the ROSAT/PSPC and corrected for vignetting.
The contour levels are logarithmically spaced with the peak
brightness corresponding to $9.4 \times 10^{-2}\,{\rm cts\,arcmin}^{-2}$.
Superimposed are the SuZIE drift scans made in 1994; one row
 of detectors
passes over the X-ray center and the other $2.2\pr$ to the south.
Scans were alternately begun $12\pr$ (shaded) and $18\pr$ ahead of the
X-ray cluster.}
\label{Obscheme}
\end{figure}

The ROSAT/PSPC data were used to determine the
spatial dependence of the X-ray emission. 
In Fig.~\ref{Obscheme}, we show the X-ray surface brightness as determined by the PSPC.
The peak surface brightness was found at 
$16\ah\,15\am\,46\as,\;-06^{\circ}\,08\pr\,55\2pr\;({\rm J}2000)$.
The radial profile of the surface brightness (corrected for vignetting)
was found by summing annuli about this central value,
with significant emission detected up to $18\pr$.
The surface brightness profile was fit with the combination of 
an isothermal $\b$ model (equation~\ref{neq}) 
convolved with the instrument point spread function (PSF) 
and a constant background.
The best fit density model parameters were found to be
$\b = .62_{-.02}^{+.015}$
and $\theta_c = 1.20 \pm 0.075\pr$ at $90\%$ confidence (\cite{Elbaz}).

Using the ASCA satellite it is, in principle, possible to measure large-scale
variations in the temperature of the IC gas.
Combining the ASCA/SIS+GIS spectra with the ROSAT/PSPC derived density
model, Markevitch \ea (1996) 
determined the temperature of the
IC gas in three radial bins of $0-3$, $3-6$, and $6-13$ X-ray core radii
to be $12.2^{+1.9}_{-1.2}\,$keV, $11.5^{+2.7}_{-2.9}\,$keV, 
and $3.8^{+1.1}_{-0.9}\,$keV at $90\%$ confidence.

As noted by Markevitch \ea (1996), the
ASCA determined
temperature is systematically lower than that found by GINGA. The 
emission-weighted ASCA/GIS+SIS average temperature for the combination of
the 3 radial bins is
${\bar T}_e = 11.2 \pm 1.1 \,$keV at $90\%$ confidence (Markevitch,
private communication), while an analysis of the GINGA data gives 
${\bar T_e} =14.6^{+.9}_{-.8}\,$keV, also at $90\%$ confidence 
(\cite{Elbaz}).
These two temperature determinations are inconsistent at $99.99\%$
confidence. It is beyond the scope of this paper to try to
understand the origin of this discrepancy, part of which may
originate in the calibration of the instrument effective areas.

In Section~\ref{ctemp}, we estimate the uncertainty in the central and 
emission-weighted average temperatures of the IC gas taking this systematic 
difference into account.
In addition, we discuss the effect of the uncertainty in temperature 
on the other X-ray parameters relevant for the determination of the 
Hubble constant.
In Section~\ref{ASCAts}, we consider the ASCA
determined thermal structure and introduce a simple two parameter model
for the IC gas temperature profile. 
In Section~\ref{sbts_hyb}, we repeat the
analysis of the ROSAT imaging data consistent with 
each of the models for the large scale thermal structure. 
We also determine the uncertainties on the X-ray
normalization and density profile parameters due to 
systematic errors in the background and uncertainties in the cluster
extent, which were not considered in \cite{Elbaz}.

\subsection{Temperature Profile}
\label{ASCAts}
Independent constraints on the IC gas thermal structure can
be derived from the combination of the X-ray derived 
density distribution and S-Z surface brightness. 
To facilitate this, we parameterize the thermal structure in terms of a
hybrid model for
the radial temperature profile of the IC gas.
The model consists of an isothermal central region extending to 
$\theta_{iso}$ beyond which the
temperature decreases according to a polytropic model with index $\gamma$,
\begin{equation}
\begin{array}{l}
        T_e(\theta) = \left\{
        \begin{array}{ll}
                T_{iso} & \theta \leq \theta_{iso} \\
                T_{iso}\left[\frac{1+\left(\theta/\theta_c\right)^2}{1+\left(\theta_{iso}/\theta_c\right)^2}\right]^{-\frac{3}{2}
                \b \left(\gamma -1 \right)} & \theta > \theta_{iso}\,.
        \end{array}  
        \right.   
\end{array}   
\label{Thybrid}
\end{equation} 
This model was originally introduced by Hughes \ea (1988) 
in their study of the Coma cluster which was, prior to ASCA, the
only cluster for which the radial temperature profile had been measured at
large distances from the cluster center. In Coma, the IC gas temperature 
decreases significantly at large radii.
In the analysis of
EXOSAT, Tenma, and GINGA data for the Coma cluster,
Hughes \ea (1993) set $\gamma = 1.55$ and determined a
best fit isothermal angular radius $\theta_{iso} \approx 3.5\,\theta_{c}$.
We have generalized this model by allowing both $\gamma$ and $\theta_{iso}$ to
vary.

\begin{figure}[htb]
\plotone{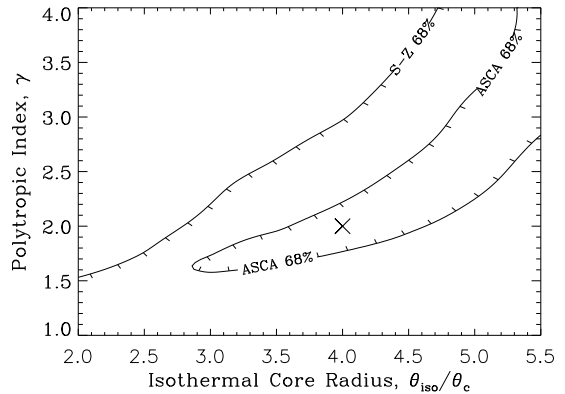}
\caption[]
{Fits of the ASCA and S-Z data to the hybrid thermal models.
The hatched regions
contain the allowed values of the model parameters at $68\%$ confidence.
The ROSAT fits constrain $\gamma < 2.5$, especially for small 
$\theta_{iso}$ .
The ``X'' marks the adopted hybrid model.}
\label{szasca}
\end{figure}

This parametric model provides a convenient analytical description of
the temperature profile measured by ASCA. 
In Fig.~\ref{szasca}, we plot the approximate
$68\%$ confidence interval for the hybrid model parameters 
($\gamma,\,\theta_{iso}$) allowed by the ASCA data. 
These parameters are constrained by computing
the emission-weighted average temperature in the three ASCA radial bins
for each of the hybrid models considered. We determine the $\chi^2$
for the fit of the models to the measured ASCA/GIS+SIS temperatures,
leaving the central temperature as a free parameter. 
A range of model parameters for the thermal structure provide a good fit 
to the ASCA data; we adopt $\theta_{iso}=4.0\,\theta_c$ and $\gamma = 2.0$ 
as representative values.

\subsection{Central and Average Temperature}
\label{ctemp}
The discrepancy between the ASCA and GINGA determined temperatures
makes the comparison of the isothermal and non-isothermal
treatments difficult.
In this section, we attempt to remedy this situation by determining a 
consistent X-ray temperature from the two experiments.

Markevitch \ea (1996) simultaneously fit the 
ASCA/SIS+GIS and GINGA spectra to determine $13.3$, $13.3$, and $3.8\,$keV 
in the three radial bins defined in Section~\ref{Xanal}. 
For the analysis including the thermal structure, we
adopt $T_{e0}=13.3\,$keV as the central IC gas temperature.
We determine the uncertainty of the central temperature taking into account
systematic difference between the two experiments. 
Crudely scaling the error bars of the ASCA/GIS+SIS central 
temperature to $68\%$ confidence, we find 
$T_{e0} \approx 12.2^{+1.2}_{-.7}\,$keV.
Assuming the thermal structure determined by ASCA,
the central temperature is $\sim 7\%$ higher than the isothermal
value.  
In this case, the central temperature determined by GINGA increases to 
$T_{e0} \approx 15.6^{+.6}_{-.6}\,$keV. 
We adopt error bars for the combined GINGA and ASCA temperatures that 
enclose the $68\%$ confidence intervals from each of the experiments 
considered individually:
ASCA $(T_{e0}\greatsim 12.2-0.7)$ and GINGA 
$(T_{e0}\lesssim 15.6+0.6)$. 
Therefore, assuming the ASCA determined thermal structure,
we find $T_{e0} \approx 13.3^{+2.9}_{-1.8}\,$keV at $68\%$ confidence.
In Section~\ref{HTS}, we use this value of the temperature  
to determine $H_0$ under the assumption of the 
ASCA thermal structure.

We also determine a consistent temperature for the case in 
which the IC gas is assumed to be isothermal.
Using the ASCA+GINGA central temperature and the
the ASCA/GIS+SIS thermal gradient, we estimate the emission-weighted 
average temperature to be ${\bar T}_e \approx 12.4\,$keV. 
The ASCA/GIS+SIS emission-weighted temperature is
${\bar T}_e \approx 11.2 \pm 0.7\,$keV at $68\%$ confidence.
We adopt error bars for the combined GINGA and ASCA analysis that 
enclose the $68\%$ confidence intervals from 
both the ASCA $({\bar T}_e \greatsim 11.2-.7\,{\rm keV})$ and GINGA 
$({\bar T}_e< 14.6+0.6\,{\rm keV})$ isothermal spectral analysis. 
Therefore, assuming the IC gas to be isothermal, we find
${\bar T}_e \approx 12.4^{+2.8}_{-1.9}\,$keV at $68\%$ confidence.
In Section~\ref{hisogas}, we use this value of the temperature to 
determine $H_0$ assuming the IC gas to be isothermal.

\subsubsection{X-ray Emissivity}
\label{lamda}
The temperature uncertainty contributes to the uncertainty of the central
specific emissivity ($\L_{e0}$) in equation~\ref{ANG DIA}.
$\L_{e0}$ is obtained by convolving $\L(T_e,E)$ with the ROSAT/PSPC
response (\cite{Elbaz}).
The emissivity  varies by $\D \L_{e0}/\L_{e0}= ^{+2.5}_{-4.0}\%$ for $T_{e0}
= 13.3^{+2.9}_{-1.8}\,$keV; similar errors are obtained for an isothermal
plasma with ${\bar T}_e  = 12.4^{+2.8}_{-1.9}\,$keV.

We have fixed $N_{\rm H}$ and the  metallicity to their best fit values as 
determined by \cite{Elbaz} (as was done by \cite{Markevitch96} in their
ASCA analysis). However, there are uncertainties in these values; from
the ROSAT/GINGA analysis, ${\rm log}(N_{\rm H}) = 21.22 \pm 0.03$ and the
metallicity is $0.4 \pm 0.1$ at $68\%$ confidence. The
corresponding uncertainties on $\L_{e0}$ are $\pm 2.5\%$ and $\pm1\%$,
respectively. We have assumed that these additional uncertainties are
independent from the uncertainty in the temperature. In particular, the 
$N_{\rm H}$ value is constrained by the PSPC spectrum which is
insensitive to the exact temperature.

Adding the uncertainties in quadrature, we obtain a statistical uncertainty in the
central specific emissivity, $\D \L_{e0}/\L_{e0}=^{+3.7}_{-4.8}\%$. 
This uncertainty is 
correlated with the temperature uncertainty ($\L_{e0}$ decreases with
increasing temperature) and is included in the uncertainty in $H_0$
due to variations in temperature.

\subsection{X-ray Surface Brightness Analysis}
\label{sbts}
\subsubsection{Isothermal Model}
\label{sbts_iso}
The normalizations ($N_X$, $N_{SZ}$) and the density model parameters
($\b,\, \theta_c$) are correlated. 
The term, $N_{SZ}^2/N_X$, which appears in equation~\ref{ANG DIA}, must be 
determined consistently from the X-ray and S-Z data.
The statistical precision and
spatial resolution of the X-ray imaging data for A2163 are much better than
the corresponding precision in the S-Z data. As a result, the density
profile shape is constrained by the X-ray surface brightness
profile. 
The error on the peak Comptonization  $y_0$, determined
from the S-Z data, is dominated by the statistical errors on the S-Z
data and not by the uncertainties in the density model parameters. 
We therefore 
express $N_{SZ}^2/N_X$  in terms of the nearly independent quantities
$N_X^{\pr}$ (determined from X-ray imaging data only) and $y_0$, 
\begin{equation}
\frac{N_{SZ}^2}{N_X} =
\frac{y_{0}^2}{N_X^{\pr} }\,,
\end{equation}
where
\begin{equation}
 N_{X}^{\pr} = b_{X0} \frac{ \Theta_{SZ0}^2}{\Theta_{X0}} = b_{X0}\,
 \theta_c
\frac{B^2\left(\frac{1}{2}, \frac{3}{2}\b - \frac{1}{2}\right)}{
B\left(\frac{1}{2}, 3\b - \frac{1}{2}\right)}\,,
\label{Xnorm}
\end{equation}
$b_{X0}$ is the central X-ray surface brightness, and $B$ is the incomplete 
beta function.

Because $b_{X0}$ and the density model parameters are
not independent, we directly compute the uncertainty in $N_{X}^{\pr}$ from fits
to the original PSPC surface brightness profile obtained by 
Elbaz \ea (1995).
A $\chi^2$ procedure, as used in Avni (1976), is applied to 
determine the uncertainty with the other parameters ($\b$,
$\theta_c$, and background level) being optimized for any given value of
$N_{X}^{\pr}$. We find the statistical uncertainty to be  $\D
N_X^{\pr}/N_X^{\pr}=\pm 3.5\%$ at $68\%$ confidence. The
absolute calibration of the PSPC is uncertain by  $\sim \pm 5\%$.
Adding the uncertainties in quadrature, we determine the X-ray
normalization to be uncertain by $\D N_X^{\pr}/N_X^{\pr} = \pm 6.1\%$.
At $68\%$ confidence, the density model parameters are $\b
= .616_{-.009}^{+.012}$ and $\theta_c = 1.20 \pm 0.05\pr$.

The derived density profile shape depends on the background level.
The background level is a free parameter in the fit to the surface 
brightness profile and the  statistical errors in
the background are included in the uncertainties of the density model 
parameters. 
In the above analysis, variations of the background with position due 
to improper vignetting corrections and contributions from unresolved 
point sources (especially at the outer part of the field of view) 
are not taken into account.
As discussed in \cite{Elbaz},
these background variations are $\sim 9\%$.
Squires \ea (1996) 
determined the effect of the background uncertainty on the determination of
the density model by repeating the fits with the background $\pm$10\% 
from the nominal value.
The additional uncertainties
in the model parameters are $\Delta\b = \pm 0.03$ and $\Delta\theta_c =
\pm 0.1\pr$. 
The corresponding additional uncertainty on the structural form factor
($\Theta_{SZ0}^2 / \Theta_{X0}$) is $\pm 2.5\%$. Adding
these systematic errors in quadrature with the statistical errors, we
obtain $\b = .616\pm 0.031$, $\theta_c = 1.20 \pm .11\pr$, and $\D
N_X^{\pr}/N_X^{\pr} = \pm 6.6\%$. 

\subsubsection{Truncated Model}
\label{trunk}
The isothermal $\b$ model (equation~\ref{neq}) is a good fit to the 
X-ray data within the radius of
maximum detection ($\theta_{max} = 15\,\theta_c$),
at which the signal to noise ratio drops to unity (\cite{Elbaz}).
For values of $\b < 1.0$, this simple model implies an
infinite mass of IC gas. Therefore, at large radius, the density
distribution of A2163 ($\b = .616$) must decrease more rapidly than
the inferred density profile.
One way to obtain a more realistic distribution
is to assume that the gas density falls to zero at a 
cut-off radius ($\theta_{cut}$) where $\theta_{cut} \geq \theta_{max}$.
We have fit the X-ray data with models for the surface brightness due
to a truncated density distribution for a range of cut-off radii,
$\theta_{max} \leq \theta_{cut} \leq \infty$.

In the most extreme case ($\theta_{cut}=\theta_{max}$), we find
$\b = .625$ and $\theta_c=1.22\pr$. 
These values are both within the $68\%$ confidence intervals
determined using the standard $\b$ model.
The corresponding change in the X-ray normalization is
$N_X(\theta_{cut}=15\,\theta_c)/N_X(\infty) = .98$. 

\subsubsection{Hybrid Model}
\label{sbts_hyb}
In the presence of thermal structure, the observed morphology of the
X-ray surface brightness will be a function of both the temperature
of the IC gas and the energy band over which the measurement is made.
We repeat the analysis of the ROSAT/PSPC data assuming
models for the thermal structure given by equation~\ref{Thybrid}. 

We first compute the specific emissivity by convolving 
$\L(T_e,E)$ with the
ROSAT/PSPC response in the considered energy band. A redshifted isothermal
plasma model is used, where the hydrogen column density and the heavy metal
abundance are fixed to their best fit isothermal values. Assuming values
for the density and thermal model parameters, we then create a model for
the X-ray surface brightness by numerically integrating equation~\ref{X-surf}.
Each model, after convolution with the PSPC PSF and
the addition of a constant background, is fit to the observed surface
brightness profile. For a given thermal model, we then determine the best
fit density model parameters and X-ray
normalization, by minimizing the $\chi^2$ of the model fit.  We have done
this for a grid of hybrid thermal models described by the parameters
$(\theta_{iso},\,\gamma)$. 
For the adopted thermal model ($\gamma=2.0$ and 
$\theta_{iso}=4.0\, \theta_c$),
we determine $\beta=0.640$ and $\theta_c=1.26\pr$.
These values are listed, along with the isothermal model
parameters in Table~\ref{partab}.
In Section~\ref{HTS}, each model for the thermal
structure is combined with its corresponding ROSAT/PSPC derived density
profile in order to determine a consistent model of the S-Z surface
brightness. 

\renewcommand{\arraystretch}{1.25}
\begin{table*}[htb]
\begin{center}
\begin{tabular}{cccccc}
\multicolumn{6}{c}{Temperature and Density Model Parameters}\\\tableline\tableline
Thermal Structure & $kT_{e0}\,$[keV] & $\gamma$ & $\theta_{iso}[\theta_c]$ & $\beta$ &
$\theta_c$\\\tableline
Isothermal & $12.4^{+2.8}_{-1.9}$ & $1.0$ & $-$ & $0.616$ & $1.20$\\
Truncated &  $12.4^{+2.8}_{-1.9}$ & $\infty$ & $15.0$ & $0.625$ & $1.22$\\
Hybrid(ASCA) & $13.3^{+2.9}_{-1.8}$ & $2.0$ & $4.0$ & $0.640$ & $1.26$ \\
\end{tabular}
\end{center}
\caption{Temperature and density model parameters for the
isothermal, truncated, and hybrid models.}
\label{partab}
\end{table*}
\renewcommand{\arraystretch}{1.0}

The changes in the best fit density profile parameters are,
for  most models considered, smaller than the corresponding
statistical uncertainties. However, for all models with 
values of $\gamma > 2.5$, especially those with small $\theta_{iso}$,
the fit to the PSPC surface brightness is bad. 
The temperature reaches such a low value that the X-ray emission in 
the PSPC band cuts off sooner than the observed surface brightness
distribution. 
This constrains the acceptable temperature model parameters to lie 
within the range plotted in Fig.~\ref{szasca}.

\section{S-Z Observations}
\label{mmobs}
\subsection{Instrument}
\label{Instrument}
All S-Z observations were made using the Sunyaev-Zel'dovich
Infrared Experiment (SuZIE) bolometer array at the Caltech
Submillimeter Observatory (CSO) on Mauna Kea.
A detailed description
of the SuZIE instrument has been presented elsewhere (\cite{Holzapfel96a}).
SuZIE is a $2 \times 3$ array of $300\,$mK bolometric detectors optimized
for the
observation of the S-Z effect in distant $(z > .1)$  clusters of galaxies.
A tertiary mirror re-images the Cassegrain focus of the
telescope to an array of parabolic concentrators which couple the
radiation to the six individual composite bolometers (\cite{Alsop}).
In order to reduce spill-over, a $2\,$K stop limits the illumination
of the CSO $10.4\,$m primary mirror to $50\%$ of its total area.
The array consists of two rows separated by $2.2\pr$;
each row consists of three co-linear array elements separated by 
$2.3\pr$.
Each array element produces a beam on the sky which is approximately
$1.75\pr$ FWHM.

The spectral response of the array elements is determined by a
common set of metal-mesh filters positioned between the parabolic
concentrators and the cold stop.
The filters can be changed for observations in several mm-wavelength
passbands.
The $2.1\,$mm filter passband is designed to maximize the ratio of the
S-Z signal to the sum of the atmosphere
and detector noise and has a FWHM $\D\lambda/\lambda=.11$.

Array elements within each of the two rows are electronically
differenced by placing pairs of detectors in AC biased bridge circuits.
The output of each bridge is synchronously demodulated to produce a
stable DC voltage proportional to the instantaneous brightness
difference between the two array elements.
The electrical bias on each detector is
adjusted so that the responsivities of the three detectors in each
row are matched (\cite{Glezer}).
The bolometer differences strongly reject signals common to both elements,
due to fluctuations in the temperature of the $300\,$mK heat sink
and atmospheric emission.
Each row produces three differences, two with beams separated by 
$2.3\pr$ and one with a $4.6\pr$ separation.
During all observations, the telescope is
fixed in place and the rotation of the earth drifts the source across
the array of detectors.
Keeping the telescope fixed while taking data
eliminates signals due to modulation of the
telescope's sidelobes and microphonic response of the detectors.
Between scans, the array is rotated about the optical
axis so that the two rows of the array are kept parallel to the
direction of the scan.
The difference signal from each pair of detectors and the absolute
voltage on each of the detectors is low-pass filtered at $2.25\,$Hz
and digitally sampled at $5\,$Hz.

\subsection{Observations}
\label{Observations}
A2163 was first observed on the nights of April 23-26, 1993 for a total
of 16 hours of integration at $2.1\,$mm;
these observations have been described in W94.
W94 did not include the
data from the night of April 23 in their analysis due to the
short observation time and comparatively poor quality of the data.
We include these results in this reanalysis, however, they
make almost no contribution to the weighted mean.
We observed A2163 at $2.1\,$mm on the nights of 1994 April 4, 9, 10
and 11 for a total of 8 hours.
The array was positioned so one row passed over
the X-ray center as determined by the ROSAT/PSPC (\cite{Elbaz}),
while the second row passed $2.2\pr$ in declination to the south.
Scans were $30\pr$ long and begun at a Right Ascension
Offset (RAO) leading the source by either $12\pr$ or $18\pr$.
In Fig.~\ref{Obscheme}, we show a schematic of the 1994 scan strategy
overlaid on the measured X-ray surface brightness contours.
First, the telescope is pointed to a position leading
the source by one of the two fixed RAOs.
The telescope tracks this position while the data
from the last scan is compressed and stored, the reciever is rotated, and
the computers perform housekeeping chores.
After about $10\,$s, the telescope stops tracking and a new scan begins;
while gathering data, the telescope is held fixed.
At the end of the scan, the telescope is reset to the alternate RAO and
the cycle is repeated.
As is described in Section~\ref{baseline}, alternating the RAO between
sequential scans provides a sensitive test for instrumental baselines.

Over the course of the observations, the zenith angle of A2163
ranged from $50^{\circ}$ to $57^{\circ}$.
The optical depth of the atmosphere at the CSO was simultaneously
measured with a tipping radiometer operating at $225\,$GHz.
Converted to the frequency and zenith angle of these measurements,
the atmospheric absorption ranged from $2-4\%$ over the course of
the observations and has been corrected for in the calibration of
the data.

As a test for a possible instrumental baseline we observed 
regions free of known sources, 
selected so they could be observed over a similar range of
hour angle and zenith angle as A2163.
The 1993 observations were coupled with $6.7$ hours of integration on a 
patch of sky  centered at  
$13\ah\,11\am\,29\as,\;-1^{\circ}\,20\pr\,11\2pr\;({\rm J}2000)$.
In 1994, we accumulated 8.5 hours of data on patch \#1 centered at
$10\ah\,24\am\,25\as,\;3^{\circ}\,49\pr\,12\2pr\;({\rm J}2000)$,
and 7.5 hours on patch \#2 centered at
$16\ah\,32\am\,44\as,\;5^{\circ}\,49\pr\,41\2pr\;({\rm J}2000)$.

\subsection{Calibration}
\label{cal} 
In April 1993, scans of Jupiter, Saturn, Mars, and Uranus were made
in order to map the beam shapes of the instrument and calibrate
its responsivity.
In W94, Mars was used as the primary
calibration source.
The scans of Mars were made with the receiver rotated by an angle
somewhat larger than that used in the source observations.
We have reanalyzed the April 1993 observations using Uranus for
the calibration.
The scans across Uranus were made with a receiver rotation angle of
$10^{\circ}$ which this lies near the center of the range of rotation angles
under which A2163 was observed. 
This reduces the uncertainty
in calibration due to changes in beam shape with receiver rotation.

In April 1994, both Uranus and Jupiter were available as calibration
sources.    
As in the reanalysis of the 1993 data, Uranus was used both to map
the beams and as the primary responsivity calibrator.
The beam-shapes measured with Jupiter and Uranus exhibited a small and
reproducible dependence on the rotation angle of the instrument about the
optical axis.
Over the course of the A2163 observations, the rotation
angle varied from $-45^{\circ}$ to $+37^{\circ}$.
The calibrated models computed for the range of rotation angles differ 
by $< 5\%$.

In order to calibrate the data, we determine the ratio of
the S-Z brightness to the
brightness of the calibrator.
The brightness of Uranus is found from a third order polynomial
fit to the measured brightness temperature as a function of
wavelength (\cite{Griffin}).
They assign an uncertainty of $\pm 6\%$ to the brightness of Uranus,
most of which arises from the $\pm 5\%$ uncertainty in the
absolute brightness of Mars from which Uranus is calibrated.
The spectral response of the array elements,
including detailed checks for out of band leaks,
have been measured using a Fourier transform spectrometer.
The central frequencies of the six array elements are
determined with an absolute accuracy of better than $1\%$ and
exhibit a reproducible scatter of less than $1\%$ about the mean
(\cite{Holzapfel96a}).
Including uncertainties in the spectral calibration,
measured beam-shapes and absolute brightness of Uranus, the
uncertainty in the absolute calibration of the instrument responsivity
to the thermal S-Z effect is estimated to be $\pm 8\%$.

\section{2.1mm Analysis}
\label{2.1anal}

\subsection{S-Z Data Set}
Each of the two rows consists of three detectors: $s_1$, $s_2$, and $s_3$.
Detector signals are differenced in pairs (in hardware) to form three
difference signals: $d_{12}=s_1-s_2$, $d_{23}=s_2-s_3$, and
$d_{31}=s_3-s_1$.
These differences correspond to angular chops of $2.3^{\prime}$,
$2.3^{\prime}$, and $4.6^{\prime}$ respectively.
Each detector contributes to two of the difference signals, therefore, the
three detector differences are not independent.
To remove the degeneracy, the two $2.3^{\prime}$ differences are differenced
to form a triple beam chop,
\begin{equation}
t_{123} = d_{12}-d_{23}= s_1 - 2s_2 +s_3 \,.
\end{equation}
This combination of detector signals has the added benefit of being
insensitive to linear changes in brightness across the array,
and therefore provides superior sensitivity in the presence of
varying atmospheric emission.
The data set consists of $4$ difference signals: $d_{31}$, $t_{123}$,
$d_{64}$, and $t_{456}$ corresponding to the $4.6^{\prime}$ difference 
and triple beam chop for each of the two rows of detectors.
From now on, we refer to these four difference signals as 
$d_k$, where $k$ ranges from 1 to 4.
The average signal of the undifferenced detectors is also computed,
\begin{equation}
s=\frac{1}{6}\sum_k^6 s_{k}\,.
\end{equation}
We use the average single detector signal as a monitor of the absolute
atmospheric emission and to correct any residual common-mode response
of the detector differences.

We clean the raw data of transients due to the
interaction of cosmic rays with the detectors.
The bolometer and electronic time constants are fast enough that the
system recovers within $< 1\,$s.
These transients are idetified using an algorithm that computes the
derivative of the data in the scan and then looks for the large positive
and negative slopes associated with a cosmic ray event.
Less than $5\%$ of the data are identified as contaminated by cosmic rays.
The raw data are then binned into $3\,$s bins corresponding to $15$
$5\,$Hz samples or $0.75^{\prime}\,{\rm cos} \delta$ on the sky.
Samples flagged as bad due to cosmic rays are left out when the
bin averages are computed; bins with less than 8 samples
are not included in the analysis.

\subsection{S-Z Surface Brightness Model}
\label{mod}
The computed S-Z surface brightness profile of A2163 is $\sim 5\pr$ FWHM,
greater than the largest beam-throw of SuZIE, which is $4.6\pr$.
In order to accurately determine the central surface brightness,
we must simulate the observation of an extended source.

A model for the surface brightness of the S-Z thermal
component is computed from the X-ray surface brightness determined
density profile and the assumed thermal structure.
\begin{equation}
I_{SZ\nu}(\theta, \phi) \propto \int f_n f_{\Psi}(x,T_e) d\xi
\label{sbI}
\end{equation}
We express the surface brightness morphology in terms of a dimensionless
form factor,
\begin{equation}
S_i(\theta, \phi)= \frac{I_{SZ\nu_{i}}(\theta,\,\phi)}{I_{SZ\nu_{i}}(0,0)} = 
\frac {\int f_n(\theta,\phi,\xi)\, f_{\Psi i}(\theta,\phi,\xi) \,d\xi} {\int
f_n(0,0,\xi)\, f_{\Psi i}(0,0,\xi)\, d\xi}\,,
\end{equation}
where $S_i$ (in order to take relativistic corrections into account) is 
evaluated at the S-Z intensity weighted band center for each array element. 
The surface brightness model ($S_i$) is convolved
with the single detector beam-maps constructed from scans over planets ($V_{Pi}$)
to generate a model for the response of the detector signals,
\begin{eqnarray}
\lefteqn{sm_{K,Ti}(\theta)=}\\
& & \hspace{-20 pt} \frac{\D I_{SZi}}{I_{Pi}} \frac{1}{\Omega_P} \int V_{Pi}
(\theta-\theta^{\prime},\,\phi^{\prime}) \,
S_{K,T}(\theta^{\prime},\,\phi^{\prime})\, d\theta^{\prime} d\phi^{\prime}\,
, \nonumber
\label{smod}
\end{eqnarray}
where $\Omega_P$ is the solid angle subtended by the planetary
calibrator.
The ratio of the relativistically correct S-Z brightness to the
brightness of the planet is
\begin{equation}
\frac{I_{SZi}}{I_{Pi}} = \frac{\int f_i(\nu) I_{SZ\nu}\, d\nu}
{\int f_i(\nu) I_{P\nu}\, d\nu} \,,
\label{irat}
\end{equation}
where $f_i(\nu)$ is the measured spectral response of each array element and
$I_{P\nu}$ is the intensity of the planetary calibrator.

The models for the response of the array elements are differenced to create 
models for the response of the detector differences.
The $\sim .05^{\prime}$ resolution differential source model is 
binned identically
to the scan data to determine the model signal for each of 
the $\sim .75^{\prime}$ data bins.
The source models are then designated as $m_{ki}(RA)$, where $k$ is 
one of the four detector differences, $i$ is the position (by 
bin number) in the scan, and RA is a ofset of the model from the
nominal X-ray determined position.

\subsection{Peak Comptonization and Uncertainties: Isothermal
Model}
\subsubsection{Coadded Data Analysis}
\label{Coadd}
We coadd the difference signals over many scans to create a high
sensitivity scan of the differential surface brightness as a function of
RA.
For each difference signal (k) and scan (j), we clean the data by removing 
the best fit linear baseline,
\begin{equation}
x_{kji} = d_{kji} -a_{kj} -i\,b_{kj}\,.
\label{clean}
\end{equation}

The value of each coadded bin
i ($x_{ki}$) is given by the weighted sum of its values 
in each of $N_s$ scans,
\begin{equation}
x_{ki}= \frac{\sum\limits^{N_s}_{j=1} \frac{x_{kji}}{{\rm RMS}_{kj}^2}}{\sum\limits^{N_s}_{j=1} \frac{1}{{\rm RMS}_{kj}^2}}\,.
\label{coadd} 
\end{equation}
Each bin is weighted by the residual RMS of the scan with
the best fit (minimum RMS) linear baseline removed,
\begin{equation}
{\rm RMS}_{kj}^2 = \frac{\sum\limits_{i=1}^{N_{b}}x^2_{kji}}
{\left( N_{b} -1 \right)}\,,
\label{cowt}
\end{equation}
where $N_b$ is the number of bins in a scan.
The uncertainty in the value of each bin is determined from the
weighted dispersion of that bin for each scan about the 
mean of $N_s$ scans,
\begin{equation}
\sigma_{ki}= \sqrt{\frac{\sum\limits^{N_s}_{j=1}{\frac{(x_{ki} -x_{kji})^2}{{\rm RMS}_{kj}^2}}}{(N_{s}-1)\sum\limits_{j=1}^{N_s} \frac{1}{{\rm RMS}_{kj}^2}}}\,.
\label{coerr} 
\end{equation}

The coadded scans are fit with the isothermal source model determined in 
Section~\ref{mod}.
The best fit source amplitude ($y_0$) and position ($RA$) are found by minimizing the
$\chi^2$ of the fit to all four difference signals
to the appropriate differential source models,
\begin{equation}
\chi^2 = \sum\limits_{k=1}^{4}\sum\limits_{i=1}^{N_b}\frac{\left(x_{ki}-y_0\,m_{ki}(RA)-a_{k}-i\,b_{k}\right)^2}{\sigma_{ki}^2}\,,
\label{cafit} 
\end{equation}
where the sum over $k$ is for all the difference signals.
This procedure accurately determines the source amplitude, so long as the signal 
is a small contribution to the structure in the individual scans.
If the source contributes significantly to the structure in a single 
scan, a scan with less signal will have lower RMS and will be 
weighted higher in the coadd.
Coadding the scans using equations~\ref{clean}--\ref{cowt}
could introduce a bias in the amplitude of the coadded scans.
However, the position determined from the coadded data should be accurate.
Once the source position is known, we can repeat the analysis in a way
that takes the contribution of the source to the scan weighting into 
account.

\begin{figure}[htb]
\plotone{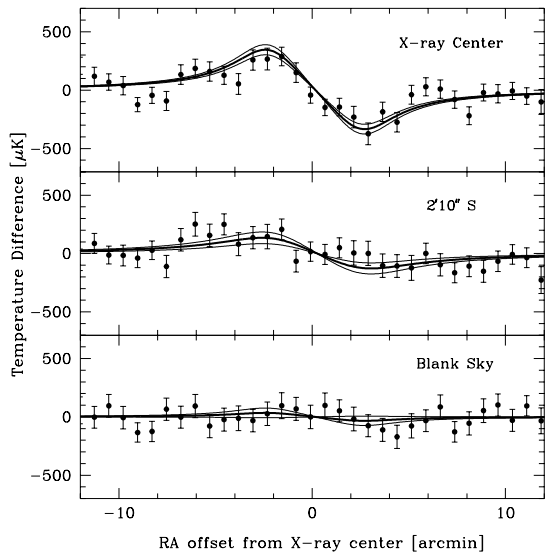}
\caption[]
{Coadded $4.6\pr$ difference data for scans across the X-ray center of 
A2163 ,
$2.2\pr$ to the south, and region \#2 of the 1994 ``blank sky''.
The data is plotted in terms of a Rayleigh-Jeans temperature
difference completely filling one of the beams.
The heavy curve is the best fit isothermal model to the
$4.6^{\prime}$ and TBC data, and the light curves correspond to the
$\pm 1 \sigma$ uncertainties on the model amplitude.}
\label{cofig}
\end{figure}

The construction of the coadded scans is repeated with the bins
weighted by the residual RMS of the scans after the removal of a the
best fit source model, residual common-mode signal, and linear baseline,
\begin{eqnarray}
\lefteqn{{\rm RMS}_{kj}^2 =} \\
& & \hspace{-20 pt}\frac{\sum\limits_{i=1}^{N_{b}} \left(d_{kji} -y_{kj}\, m_{ki}(RA)-\alpha_{kj} s_{ji} -a_{kj}-i\,b_{kj}\right)^2}
{\left( N_{b} -1 \right)}\,. \nonumber
\label{coaddnomod}
\end{eqnarray}
Allowing the source model to vary, while determining the weights,
ensures that the weighting does not bias the signal amplitude.
In Appendix~\ref{RSC}, we discuss the removal of the residual common-mode 
signal and demonstrate that it has no systematic effect on the fit
results.
We then coadd the binned scans with the best fit
linear baseline and residual common-mode signal removed,
\begin{equation}
x_{kji} = d_{kji} -\alpha_{kj} s_{ji} -a_{kj}-i\,b_{kj}\,.
\end{equation}
Using equations~\ref{coadd} and \ref{coerr}, we recompute the bin
averages and uncertainties which are free of bias due
to the non-zero source amplitude.
In Fig.~\ref{cofig}, we show coadded data scans for the 1994 
observations of A2163 and a patch of blank sky. 

The determination of accurate uncertainties for the coadded scan
fits is difficult due to the presence of atmospheric emission which
results in noise correlated between bins and detector
differences.
In Section~\ref{Scanfits}, we use the fact that
the noise from one scan to
the next is uncorrelated to determine the uncertainty in the total data
set and provide an accurate calibration of the uncertainty in the
coadded fits.

\subsubsection{Single Scan Fits}
\label{Scanfits}
Once the source position is known, each scan can be used as an 
independent measurement of the source amplitude.
The source amplitude for each scan is found by minimizing the 
$\chi^2$ of the model fit to the desired data scans,
\begin{eqnarray}
\label{chisq}
\lefteqn{{\chi^2_j} =}\\
& & \hspace{-20 pt} \sum\limits_{k}^{}\sum_{i=1}^{N_b}\frac{\left(d_{kji}-
y_{0j}\,m_{ki}(RA)- \alpha_{kj}s_{ji} -a_{kj}-i\,b_{kj}\right)^2}
{{\rm RMS}_{kj}^2}, \nonumber 
\end{eqnarray}
where RMS$_{kj}$ is given by equation~\ref{coaddnomod}. 
The uncertainty in the peak \comp\ ($\sigma_{yj}$) is simply the 
change in $y_{0j}$ corresponding to $\D \chi_j^2 =1$. 
We determine the mean peak \comp\ for a given 
observation by averaging the values $y_{0j}$ for each of the $N_s$ 
scans weighted by $\sigma_{yj}$,
\begin{equation}
y_0 = \frac{\sum\limits_{j=1}^{N_s}{\frac{y_{0j}}{{\sigma_{yj}}^2}}}{\sum\limits_{j=1}^{N_s}{\frac{1}{{\sigma_{yj}}^2}}}\,.
\label{ymean}
\end{equation}
The weighted dispersion of the scan amplitudes about the
mean is used to estimate the uncertainty in the determination of 
the mean for each observing period,
\begin{equation}
\sigma_y = \sqrt{
\frac{\sum\limits_{j=1}^{N_s}{\frac{(y_0-y_{0j})^2}{{\sigma_{yj}}^2}}}{(N_s-1)\sum\limits_{j=1}^{N_s}{\frac{1}{{\sigma_{yj}}^2}}}}\,.
\label{yerror}
\end{equation}

\begin{figure}[htb]
\plotone{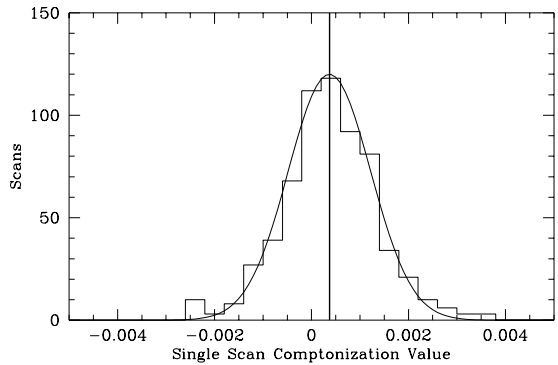}
\caption[]
{Histogram of peak \comp\ values from simultaneous fits to all four
differential signals in each individual scan.
The distribution is well approximated by the superimposed Gaussian, the
mean is marked with a vertical line.}
\label{histy}
\end{figure}

In Fig.~\ref{histy}, we show a histogram of the peak \comp\
results for simultaneous fits to both array rows for every 
A2163 scan taken in 1993 and 1994.
The mean and width of this distribution 
can be used to determine the peak \comp\ and uncertainty 
in the simple case when all scans are weighted equally.
Using this method, the mean is unchanged while the uncertainty in $y_0$ 
is $\sim 30\%$ higher than that found when the scan amplitudes are 
weighted using equation~\ref{yerror}. 
We do not use the results of this simple analysis, however, it
illustrates the independence of the results from the details 
of the analysis.

For each night, RAO, and scan declination, we use 
equations~\ref{chisq}--\ref{yerror} to compute the mean amplitude and 
uncertainty; these results are listed in Table~\ref{ty}.
The single scan amplitudes are combined  to determine an 
average peak \comp\ for all the data at each of the three scanned
declinations; 
these results are listed at the bottom of the columns in Table~\ref{ty}.
Due to correlated sky noise, the results of the 
fits to each of the two rows are not independent.
In order to estimate the uncertainty of the total data set, we
simultaneously fit to the signals from both rows.
Assuming the sky noise between scans to be uncorrelated, 
we use equations~\ref{ymean} and \ref{yerror} to determine 
$y_0=3.73 \pm .35 \times 10^{-4}$.

\begin{table*}[htb]
\begin{center}
\begin{tabular}{ccccc}
\multicolumn{5}{c}{Peak \comp\ $\times 10^{4}$ in A2163: Isothermal
Model}\\\tableline\tableline
Date &   RAO   & $\Delta\delta=0\pr$ & $\Delta\delta=+2\pr\,10\2pr$ &
$\Delta\delta=-2\pr\,10\2pr$ \\
 & & \multicolumn{3}{c}{Peak \comp, $y_0 \times 10^{4}$}\\\tableline
1993 April 23 &   $16.5\pr$ & $0.72\pm2.78$ & $-3.34\pm4.64$ & . . .\\
             & $22.5\pr$ & $2.52\pm2.15$ & $9.33\pm6.36$ & . . .\\
1993 April 24 & $16.5\pr$ & $5.06\pm1.39$ & $7.33\pm2.36$ & . . .\\
              & $22.5\pr$ & $3.40\pm1.66$ & $0.51\pm2.32$ & . . .\\
1993 April 25 & $8.5\pr$ & $4.46\pm1.17$ & $1.74\pm2.26$ & . . .\\
             & $14.5\pr$ & $3.30\pm1.22$ & $4.96\pm2.20$ & . . .\\
1993 April 26 & $8.5\pr$ & $4.39\pm0.82$ & . . . & $1.32\pm2.32$\\
            &  $14.5\pr$ & $6.71\pm1.08$ & . . . & $2.03\pm2.48$\\
1994 April 4 & $12\pr$ & $5.38\pm2.42$ &  . . . & $11.43\pm7.33$\\
             & $18\pr$ & $7.01\pm1.20$ & . . . & $15.11\pm5.59$\\
1994 April 9 & $12\pr$ & $3.20\pm0.77$ & . . . & $3.58\pm2.20$\\
            & $18\pr$ & $2.41\pm0.83$ &  . . . & $1.67\pm2.61$\\
1994 April 10 & $12\pr$ & $3.08\pm1.65$ & . . . & $-2.18\pm3.53$\\
            & $18\pr$ & $2.25\pm1.08$ & . . . & $3.17\pm3.17$\\
1994 April 11 & $12\pr$ & $4.36\pm1.04$ & . . . &  $1.05\pm3.10$\\
            & $18\pr$ & $3.54\pm1.23$ & . . . & $2.03\pm3.42$\\\tableline
\multicolumn{2}{l}{Weighted mean} & $3.87\pm0.35$ & $3.50\pm1.10$ & $2.77\pm0.97$
\end{tabular}
\end{center}
\caption{Results of fits of the isothermal model
to the data for each observation, $\delta$, and RA.
The total results for each declination (bottoms of columns)
are calculated in the same way as the individual sets using
the distribution of the scan amplitudes.}
\label{ty}
\end{table*}

The average source amplitudes determined by the single scan fits agree 
with the results of the fits to the coadded data,
while the uncertainties are a factor of $1.4$ to $2.0$ higher than 
those determined from the fits to the coadded data.
This is because the atmospheric noise, which dominates
the noise in the scans, is correlated between difference signals 
and bins in a given scan.
In order to use the coadded data to determine confidence intervals 
for the position and morphology of the S-Z surface brightness, 
we must take the correlation of the noise into account.
To do this, we scale the uncertainties of the coadded bins so
that the uncertainty in the coadded scan amplitude is the same as 
that from determined the single scan fits.
The scaling factor is insensitive to the details of the 
cluster position and morphology.
We can then use the coadded data to determine the
confidence intervals for parameters such as the model position which
cannot be determined from the individual scan fits.
All fits to coadded data discussed in this paper are for
data with bin uncertainties scaled to compensate for 
correlated noise.

\subsubsection{Peak S-Z Surface Brightness Position}
\label{szposition}
With the coadded scan bin uncertainties appropriately scaled,
we use a maximum likelihood indicator to determine confidence
intervals in the model amplitude and position.
We determine the position of the peak S-Z surface brightness
to be offset from the peak X-ray surface brightness by
$\D RA = +.35 \pm .14 \pr$ at $68\%$ confidence.
The value of the peak \comp\ decreases by less than $1.0\%$
from its value at the central position over the allowed range of RA.

We estimate the uncertainty in pointing the SuZIE instrument to be 
$\lesssim 10^{\2pr}$.
ROSAT astrometry is typically uncertain by $10-15\2pr$,
unless special care has been taken to locally align X-ray and optical
frames. 
Because no X-ray bright optical sources were present in the ROSAT field,
this was not possible for A2163.
Including the uncertainty in the ROSAT and SuZIE pointing,
the offset between the X-ray and S-Z peak surface brightnesses
is then $\D RA = +.35 \pm .33^{\prime}$ at $68\%$ confidence.

We do not have sufficient coverage to
accurately locate the peak S-Z surface brightness
in declination.
Assuming the peak S-Z and X-ray surface brightnesses to be coincident,
the $\delta$ of the peak surface brightness is uncertain
by the quadrature sum of the ROSAT and SuZIE pointing uncertainties,
$\D \delta \approx .30^{\pr}$.    
If the true position of the peak surface brightness is offset from
the declination of our scan, the measured peak \comp\ 
will be smaller than the true value.
Fitting the data to a model for scans offset in $\delta$ 
from the peak surface brightness, we find that the
peak \comp\ is $\sim 2.0\%$ smaller for $\D \delta =\pm .30^{\prime}$.
The uncertainty in the position of the peak 
S-Z surface brightness, therefore, results in an uncertainty in the 
peak \comp, $\D y_0/y_0= +2.2\%$.
This contribution to the total uncertainty is listed under
``Position'' in Table~\ref{yut}.

\subsubsection{Baseline}
\label{baseline}
We analyze the ``blank sky'' scans across regions of sky free of known sources
(described in  Section~\ref{Observations}) exactly as the source data,
to test for an instrumental baseline.
An instrumental baseline with the same temporal features as the source model 
would be detected as a non-zero \comp\ for these regions. 

In the April 1993 observations, the RAO of sequential scans was 
alternated in the same way as the for the source data. 
In W94, this data was used to estimate the baseline 
contribution to the uncertainty in the source amplitude. 
Reanalyzing this data using the ROSAT density model and the 
residual common-mode signal subtracted scans, we find
$y_0 = -0.1 \pm 4.2 \times 10^{-5}$

In the April 1994 observations (patch \#1 and patch \#2),
the RAO was alternated between $12^{\prime}$ and $18^{\prime}$ in sequential
scans. The results of these observations have been used to place upper 
limits on arcminute scale CMB anisotropies (\cite{Church}).
Analyzing the baseline data in the same way as the source data,
we find $y_0 = 1.3 \pm 3.1 \times 10^{-5}$ and 
$y_0 = 1.8 \pm 3.1 \times 10^{-5}$
for the two patches, respectively.
In Fig.~\ref{cofig}, we show the coadded $4.6^{\prime}$ difference 
for the scans on patch \#2.
Combining the results from all three patches,
we determine $y_0 = 1.45\pm 1.9\times 10^{-5}$, indicating no significant
instrumental baseline.
Combining the results from multiple patches reduces the signal from 
potential sources of confusion but should accurately determine the 
effect of any common instrumental baseline. 
One problem with this method is that the baseline data and source 
data, although gathered at similar azimuth and zenith angle, are 
not gathered simultaneously or even on the same evening.
Any baseline signal, because it is not significant in a single scan,
must be correlated in time between several scans.
This does not guarantee the baseline will be constant
over the course of an observation.

We have devised a method which allows us to use the source
data to test for the presence of a scan-correlated baseline.
Differencing scans taken adjacent in time at
two different RAOs allows the subtraction of the common baseline
with little effect on the expected signal.
Each pair of scans are differenced by subtracting their raw 
$5\,$Hz sampled time streams.
The data are then analyzed exactly as described in 
Sections~\ref{2.1anal}--\ref{baseline}.

We construct a model for the differential surface brightness by
differencing two models with the appropriate RAO difference between them.
We choose the RAO difference ($6^{\pr}$) to be large enough so that the
two models are well separated in the scan and therefore will 
not subtract much signal when differenced.
The differenced model is then fit to the correlation corrected coadded 
scans in order to determine the source amplitude and position.
Fitting to the isothermal model, we find $y_0 = 3.49 \pm .40 \times 10^{-4}$
and $\D RA = +0.19 \pm 0.18^{\prime}$ at $68\%$ confidence.
These results agree, within the quoted statistical errors, with those 
determined from the undifferenced data.

The lack of any significant change when the scans are differenced,
as well as the null result for the ``blank sky'' fields, indicates 
that there is no significant instrumental baseline with the same
temporal dependence as the source model.
The difference between the results of the fits to both RAOs 
and the differenced scans is treated as an additional 
contribution to the uncertainty in $y_0$ and is listed
under ``Baseline'' in Table~\ref{yut}.
Including the additional uncertainty due to the possibility of an 
instrumental baseline, the results from Section~\ref{Scanfits}
become $y_0 = 3.73^{+.35}_{-.43} \times 10^{-4}$.

\subsection{Density Model Uncertainties}
\label{yDMU}
 We consider now the
effect of the uncertainties in the density model on the determination of $y_0$.
We have fit the S-Z data with the range of density models
allowed by the X-ray analysis.
The uncertainties in the density model parameters 
(Section~\ref{sbts_iso}) contribute an uncertainty in the peak
\comp\ of $\D y_0/y_0=^{+2.5}_{-3.6}\%$ at $68\%$ confidence.
In W94, the data were fit with a template derived from the
density model determined from the analysis of Einstein data:
$\theta_c = 1.15$ and $\beta = 0.59$ (\cite{Arnaud92}).
In this paper, these data are reanalyzed using the ROSAT/PSPC density model
which results in a peak \comp\ lower by $\sim 4\%$.

The truncated density model, introduced in Section~\ref{trunk},
is used to generate a model for the isothermal S-Z surface brightness. 
Fitting the measured S-Z surface brightness with the model for the 
truncated surface brightness, we find $y_0(\theta_{max})/y_0(\infty)=.925$.
The angular scale factor ($\Theta_{SZ}$) also decreases when the
density model is truncated; this nearly compensates
for the decrease in the measured peak \comp.
Even in the most extreme case, the S-Z normalization is nearly
independent of the presence of a density cut-off and
$N_{SZ}(\theta_{max})/N_{SZ}(\infty) = .985$.

The uncertainty in the peak \comp\ due to the adopted density model
is the sum of the uncertainties due to the
uncertainty in the model parameters
and the assumed functional form for the density profile.
Adding these contributions to the uncertainty in quadrature, we determine
$\D y_0/y_0 = ^{+2.5}_{-8.3}\%$.
This analysis is performed assuming an isothermal IC gas; in the
presence of thermal structure of the form discussed in 
Section~\ref{HTS}, the effect of the
cut-off density model will generally be smaller.

\subsection{Isothermal Temperature Uncertainty}
Due to our use of the relativistically correct calculation of the
\comp\ (Section~\ref{TRcomp}), the effective peak \comp\ derived from
S-Z data depends on the assumed temperature. For temperatures in the range
allowed by the GINGA and ASCA spectral
analysis (Section~\ref{ctemp}), $y_0 \propto  T_e / \Psi$ changes by 
$^{+1.6}_{-1.2}\%$. 
Including uncertainties in the baseline, calibration, and density 
distribution and temperature of the IC gas, we detrermine
$y_0 = 3.73^{+.48}_{-.61} \times 10^{-4}$.

\section{Hubble Constant: Isothermal gas}
\label{hisogas}
Using equations~\ref{ANG DIA}, \ref{yeff} and \ref{Xnorm}
we can express $d_A$ in terms of the nearly independent quantities 
$N_X^{\pr}$, $T_e$ and $y_0$,
\begin{equation}
d_A =
\frac{y_{0}^2}{N_X^{\pr} }  \left(\frac{m_ec^2}{kT_e}\right)^2
\frac{\L_{e}}{4 \pi \left(1+z\right)^3  \sigma_T^2}\,.
\label{ANG DIA2}
\end{equation}
Apart from relativistic corrections, the errors in $y_0$ and $T_e$
can be considered as independent; the corresponding contributions to the 
total uncertainty for $H_0$ are added in quadrature.

Assuming a standard $\b$ model, we determine the contribution 
of the uncertainty in the density model to the uncertainties in
$N_X^{\pr}$ (Section~\ref{sbts_iso}) and $y_0$ (Section~\ref{yDMU}).
The change in $y_0$ is small and has the same sign as the change in 
$\D N_X^{\pr}$; therefore, 
$\D H_0/H_0 \approx \D N_X^{\pr}/N_X^{\pr} = \pm 6.6\%$ 
is an overestimate of the true uncertainty. 
This contribution to the uncertainty in $H_0$ is listed in 
Table~\ref{hut} under ``X-ray normalization''. 
Compared to the contributions due to the uncertainties
in the S-Z and central temperature results, the contribution of the 
density model to the total uncertainty is nearly negligible.
Taking only the uncertainties in the S-Z and X-ray normalizations
into account, we determine
$H_0(q_0=\frac{1}{2})=59.6^{+21.5}_{-13.1}\,{\rm kms}^{-1}{\rm Mpc}^{-1}$.

In order to account for the systematic difference in the ASCA and GINGA 
temperatures,
we use the estimated isothermal {X-ray} temperature determined in 
Section~\ref{ctemp}. 
The uncertainty in $H_0$ due to $T_{e}$ includes the 
effect of the temperature uncertainty on
$\L(T_{e})$ as determined in Section~\ref{lamda}.
In Table~\ref{hts}, we list the results for $H_0$, as a function of 
the assumed thermal structure and the included sources of uncertainty. 
Including the uncertainties in the S-Z and X-ray normalizations and
X-ray temperature, we determine
$H_0(q_0=\frac{1}{2})=59.6^{+40.7}_{-22.6}\,{\rm kms}^{-1}{\rm Mpc}^{-1}$.

\renewcommand{\arraystretch}{1.25}
\begin{table*}[htb]
\begin{center}
\begin{tabular}{cccc}
\multicolumn{4}{c}{Hubble Constant (Thermal Structure)}\\\tableline\tableline
 & \multicolumn{3}{c}{$H_0\,[{\rm kms}^{-1}{\rm Mpc}^{-1}]$}\\
Thermal Structure & $N_X,\,N_{SZ}$ &$N_X,\, N_{SZ},\, T_{e0}$  & Total \\\tableline
Isothermal &  $59.6^{+21.5}_{-13.1}$ & $59.6^{+40.7}_{-22.6}$ & $59.6^{+45.3}_{-30.9}$\\
Hybrid(ASCA) & $78.4^{+31.3}_{-17.3}$ & $78.4^{+53.8}_{-27.9}$ & $78.4^{+59.9}_{-39.8}$\\
\end{tabular}
\end{center}
\caption{The Hubble constant determined for both isothermal and
hybrid thermal structure.
The results are calculated for three different cases.
In the first column under $H_0$, only uncertainties in the X-ray
and S-Z normalizations are included.
The second column includes the uncertainty in the IC gas central
temperature.
The third column includes additional uncertainties due to
peculiar velocity, astrophysical confusion, clumped IC gas, and cluster
asphericity.}
\label{hts}
\end{table*}
\renewcommand{\arraystretch}{1.0}

We also determine the effect of the truncated density model
described in Section~\ref{trunk} on the value of $H_0$.
Because $N_{SZ}$ and $N_{X}$ are insensitive to the details of the
density model, so is the value of $H_0$; for the most extreme 
case ($\theta_{cut} = \theta_{max}$),
we find $H_0(\theta_{max})/H_0(\infty) = 1.01$.
For the remainder of this paper we assume the IC gas density can be described
by a standard $\b$ model with $\theta_{cut} = \infty$.
Because the X-ray normalization contains the uncertainty due to
density model parameters, the additional contribution to $H_0$ is
simply that due to the assumption of the density functional form,
$\D H_0/H_0 = {+1.0}\%$.
This contribution to the uncertainty in $H_0$ is listed under ``Density Model''
in Table~\ref{hut}.
This analysis is performed assuming an isothermal IC gas; in the
presence of thermal structure of the form discussed in Section~\ref{HTS},
the uncertainty due to the assumption of a cut-off density
model will generally be smaller.

In Section~\ref{szposition}, we determined the effect of
the uncertainty in the position of the peak S-Z surface brightness on
the value of $y_0$.
The uncertainty in the position contributes an uncertainty 
to the Hubble constant of $\D H_0/H_0= -4.5\%$; this contribution to the total 
uncertainty is listed under ``Position'' in Table~\ref{hut}. 

\section{Large-scale Thermal Structure}
\label{HTS}
\begin{figure}[htb]
\plotone{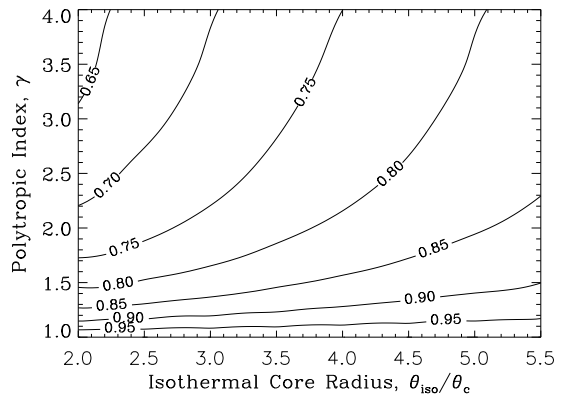}
\caption[]
{Ratio of the peak \comp\ to the isothermal value ($y_0/(y_0)_{iso}$),
plotted as a function of the hybrid thermal model parameters.}
\label{ypar}
\end{figure}

In this Section, we repeat the determination of the Hubble constant under 
the assumption of a thermal structure of the form described by 
equation~\ref{Thybrid}.
As was outlined in Section~\ref{sbts_hyb}, 
we repeat the analysis of the ROSAT surface brightness in order to 
rederive the density model parameters ($\beta,\,\theta_c$)
for each set of thermal structure parameters ($\gamma,\,\theta_{iso}$).
For the adopted hybrid model
($\gamma=2.0,\, \theta_{iso}=4.0\,\theta_c$), the reanalysis of the ROSAT
data yields density parameters $\beta=0.640$ and $\theta_c=1.26\pr$. 
Using equations~\ref{sbI}--\ref{smod} we combine the models
for the IC gas temperature and density to create models for the 
S-Z surface brightness. 
Each model is fit to the coadded data scans in order to determine
the peak \comp. 
In Fig.~\ref{szasca}, we plot the $68\%$ confidence
interval in the $(\theta_{iso},\, \gamma)$ plane for fits to
the coadded S-Z data.
The S-Z measurements lack sufficient sensitivity to detect a thermal
gradient of the type determined from recent ASCA results. 
The ratio of the peak \comp\ to that determined assuming an
isothermal gas is plotted as a function of $\theta_{iso}$ and $\gamma$ in
Fig.~\ref{ypar}.
For the range of temperature model parameters allowed by the 
ASCA results at 
$68\%$ confidence (see Fig.~\ref{szasca}), the 
values of the \comp\ span $y_0/(y_0)_{iso}=.82\pm.03$.
Using the procedure described in Section~\ref{Scanfits}, we determine the peak \comp\
from single scan fits to the hybrid S-Z surface brightness model
to be $y_0 = 3.07\pm .29 \times 10^{-4}$.
The results of the fits at each declination are listed in
Table~\ref{Tcomp}.

\begin{table*}[htb]
\begin{center}
\begin{tabular}{ccccc}
\multicolumn{5}{c}{Peak \comp, $y_0\times 10^4$}\\\tableline\tableline
Thermal Structure & $\D\delta=0\pr$ & $\D\delta=+2\pr\,10\2pr$ &
$\D\delta=-2\pr\,10\2pr$ & Total\\\tableline
Isothermal & $3.87\pm0.35$ & $3.50\pm1.10$ & $2.77\pm0.97$ &
$3.73\pm0.35$\\
Hybrid(ASCA) & $3.16\pm0.32$ & $3.34\pm1.19$ & $2.21\pm0.91$ &
$3.07\pm0.29$\\
\end{tabular}
\end{center}
\caption{Peak \comp\ at each declination for the cases when
the IC gas is either isothermal or has the ASCA determined
thermal structure.
The uncertainties include statistical uncertainty only.
The numbers under total are calculated from simultaneous
fits to both rows of the array.}
\label{Tcomp}
\end{table*}

\renewcommand{\arraystretch}{1.25}
\begin{table}[htb]
\begin{center}
\begin{tabular}{ll}
\multicolumn{2}{c}{Peak \comp\ \& Uncertainty}\\\tableline\tableline
Source & Uncertainty \\
&  $y_0\times 10^{4}$ \\\tableline
Statistical & $3.07\pm.29$ \\
Baseline & $-.27$\\
Calibration & $\pm .25$ \\
Position & $+.07$ \\
Density Model & $^{+.08}_{-.25}$ \\
Thermal Gradient (ASCA) & $\pm.08$ \\
Central Temperature & $^{+.05}_{-.04}$ \\\tableline
& $3.07^{+.40}_{-.54}$\\
\\
Peculiar Velocity & $\pm.23$ \\
Radio Confusion  & $^{+.25}_{-.05}$ \\
Primary Anisotropies & $\pm.12$ \\\tableline
\multicolumn{1}{l}{Total} & $3.07^{+.54}_{-.60}$\\
\end{tabular}
\end{center}
\caption{Peak \comp\ and contributions to uncertainty using
the adopted hybrid thermal structure.}
\label{yut}
\end{table}
\renewcommand{\arraystretch}{1.0}

\renewcommand{\arraystretch}{1.25}
\begin{table}[htb]
\begin{center}
\begin{tabular}{ll}
\multicolumn{2}{c}{Hubble Constant \& Uncertainty}\\\tableline\tableline
Source & Uncertainty \\
& \multicolumn{1}{c}{${\rm kms}^{-1}{\rm Mpc}^{-1}$}\\\tableline
X-ray Normalization & $\pm 5.2$ \\
S-Z Normalization & $^{+30.8}_{-16.5}$ \\
Position & $-3.5$ \\
Density Model & ${+.8}$\\
Thermal Gradient (ASCA) & $^{+1.5}_{-1.2}$\\
Central Temperature & $^{+43.8}_{-21.9}$ \\\tableline
 & $78.4^{+53.8}_{-27.9}$\\
\\
Peculiar Velocity & $^{+13.0}_{-10.4}$ \\
Asphericity & $\pm21.7$ \\
Clumping & $-7.8$ \\
Radio Confusion & $^{+2.6}_{-11.4}$ \\
Primary anisotropies & $^{+6.7}_{-5.9}$ \\\tableline
\multicolumn{1}{l}{Total} & $78.4^{+59.9}_{-39.8}$\\
\end{tabular}
\end{center}
\caption{Hubble Constant and contributions to the uncertainty using the
adopted hybrid thermal structure.
The S-Z normalization includes uncertainties due to statistical
uncertainty, baseline, and calibration.}
\label{hut}
\end{table}
\renewcommand{\arraystretch}{1.0}

We repeat this analysis for the RAO differenced data.
The peak \comp\ from the single scan fits is 
$y_0 = 2.80 \pm .32 \times 10^{-4}$.
We use the difference between this result and the peak \comp\
for the undifferenced scans to estimate the baseline uncertainty.
Combining the result of the fits to both RAOs with
uncertainties in the baseline, calibration, and density and temperature
models, we determine $y_0 = 3.07^{+.40}_{-.54}\times 10^{-4}$.
The density model uncertainty is assumed to be the same as in
the isothermal case (Section~\ref{hisogas}).
This is justified because the X-ray surface brightness
profile is insensitive to the thermal structure for the
considered range of thermal models.
The individual contributions to the uncertainty in $y_0$ are listed
in Table~\ref{yut}.

\begin{figure}[htb]
\plotone{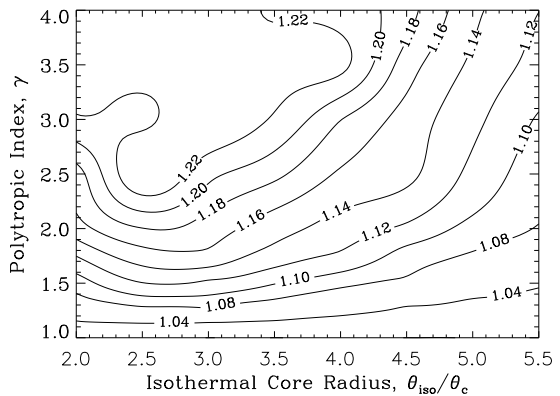}
\caption[]
{Ratio of the true Hubble constant to the isothermal value
($H_0/(H_0)_{iso}$) as a function of the hybrid thermal model
parameters.
The value of the central temperature is assumed to be constant.}
\label{hub}
\end{figure}

We combine the X-ray and S-Z results to determine $H_0$ as a function of 
the cluster thermal structure.
In this analysis, we make one of two assumptions about the results
of the X-ray observations.
In the first case, we assume that the central temperature of the IC gas 
has been determined, this requires spatial resolution in the 
spectral measurement. 
For the allowed range of thermal models, we find 
$N_X/(N_X)_{iso}= .96 \pm .02$ and $N_{SZ}/(N_{SZ})_{iso} = .92\pm{.02}$.
In Fig.~\ref{hub}, we plot the ratio of the true value of $H_0$ to that 
determined assuming the cluster gas to be isothermal at the central
temperature.
The ASCA constraints on the thermal structure (see Fig.~\ref{szasca})
limit $H_0/(H_0)_{iso} = 1.13 \pm .02$ at $68\%$ confidence.

\begin{figure}[htb]
\plotone{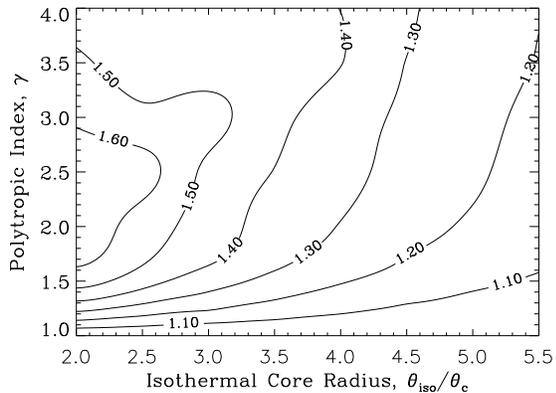}
\caption[]
{Ratio of the true Hubble constant to the isothermal value
($H_0/(H_0)_{iso}$) as a function of the hybrid thermal model
parameters.
The value of the emission-weighted average temperature
is assumed to be constant.}
\label{hubit}
\end{figure}

We also calculate $H_0$ as a function of thermal structure
for the more realistic case in which only the emission-weighted 
average X-ray temperature is known.
The values of the normalization factors ($N_X,\,N_{SZ}$) are nearly 
unchanged from the previous case.
Assuming the ASCA thermal structure, the central temperature is about
$7\%$ higher than the emission weighted average value.
This leads to a larger offset in $H_0$ from the isothermal value
than when the central temperature is held fixed.
In Figure~\ref{hubit}, we plot the ratio of the true Hubble constant 
to that calculated assuming the cluster gas to be isothermal at the 
emission-weighted average temperature.
The ASCA constraints on the thermal structure (see Fig.~\ref{szasca})
limit $H_0/(H_0)_{iso} = 1.29^{+.10}_{-.10}$ at $68\%$ confidence.
Inagaki \ea (1995) have investigated the effect of 
thermal structure on the value of $H_0$ determined from the S-Z effect.
For a simulated ``Coma-like'' cluster, they found $H_0$ to be 
underestimated by $>20\%$ when the IC gas was
assumed to be isothermal.
They identify uncertainty in the temperature profile as the most serious
systematic error in the value of $H_0$ and as a possible
source of the ``low'' values sometimes found with the S-Z effect.

We have determined $H_0$ under the assumption of the 
ASCA/GIS+SIS thermal structure modeled by
the adopted hybrid model (Table~\ref{partab}).
We find $H_0(q_0=\frac{1}{2})=78^{+54}_{-28}\,{\rm kms}^{-1}{\rm Mpc}^{-1}$
at $68\%$ confidence, where
we include uncertainties in the normalizations 
($N_X$ and $N_{SZ}$) and density and temperature models.
The uncertainties in the X-ray normalization and the density
model are considered to be the same as in the isothermal case
and all errors are added in quadrature.
The individual contributions to the uncertainty in $H_0$ are listed 
in Tables~\ref{hts} and \ref{hut}.

\section{Additional Uncertainties}
\label{Uncertainties}
The uncertainties  discussed so far are associated 
with the precision of the measurements required for the 
determination of the Hubble constant.
We identify several additional sources of systematic uncertainty 
associated with assumptions we have made in order to determine $H_0$ 
from the measured data.
Among these are deviations from spherical symmetry, peculiar velocity,
IC gas clumping, and astrophysical confusion.
The contribution of each of these additional sources of uncertainty to 
$y_0$ and $H_0$ are listed in Tables~\ref{yut} and \ref{hut}
respectively.

\subsection{Deviations From a Spherical Gas Distribution}
The assumption that the extent of the IC gas along the line of sight
is the same as is measured in the plane perpendicular to the line of sight
is fundamental to this method for determining $H_0$.
If the distribution of the IC gas is prolate (oblate)
with its dimension along the line of sight longer (shorter) by a
factor $Z$ than the average of the dimensions perpendicular to the
line of sight, then the derived $H_0$ will be
modified from the true $H_0$ by a factor $1/Z$ (BHA).
The average Hubble constant derived from measurements of N clusters,
$\bar{H_0} \propto \frac{1}{N} \sum^{N}_{n=1} \frac{1}{Z_n}$,
will be weakly biased towards higher $H_0$ depending on the distribution
of ellipticities contained in the sample.
In this paper, we quantify
the contribution of asphericity to the uncertainty in $H_0$
determined from measurements of a single cluster.
We estimate this contribution from the statistics of the ellipticity
of the observed X-ray isophotes of a sample of other clusters.

McMillian \ea (1989) determined the ellipticity of the X-ray isophotes for
a sample of 49 clusters.
The average ellipticity for this sample is $\bar{\epsilon}=.277$.
This average has been computed without the
removal of any of the clusters containing obvious substructure.
For a sample of clusters, the average elongation or shortening along
the line of sight will be smaller than the average deviation from
circularity of the projected X-ray isophotes.
Departures from sphericity of the IC gas distribution,
therefore, contribute an uncertainty of $< \pm 27.7\%$ to the
determination of $H_0$ from a single cluster.

\subsection{Peculiar Velocity}
If the IC gas has a bulk velocity (${\vec v_p}$) with respect to the
CMB rest frame, there is an additional kinematic component to
the S-Z effect,
\begin{equation}
\D I_K = -I_{0} \frac{x^4 e^x}{\left(e^x - 1\right)^2}\int{n_e \sigma_T
\frac{\vec{v_{p}}}{c} \cdot d\vec{l}} \, . 
\label{teqn}
\end{equation}
In general, the ratio of the brightnesses of the kinematic and 
thermal components of the S-Z effect integrated over the $2.1\,$mm 
band is small.
For A2163 ($T_e = 12.4 \,$keV),
\begin{equation}
\frac{\int \D I_K f(\nu) d\nu}{\int \D I_T f(\nu) d\nu} = .139
\left(\frac{v_{r}}{10^3 \;{\rm kms}^{-1}}\right) \, ,
\end{equation}
where $v_{r}=\vec{v_p} \cdot \vec{dl}$ is the projection of 
the average cluster velocity onto
the line of sight, and $f(\nu)$ is the average transmission of
the SuZIE system in the $2.1\,$mm band.

We have made additional measurements of A2163 with filter bands
centered at $1.4$ and $1.1\,$mm.
Combining the mm-wavelength measurements with $T_{e0}$
we are able to limit $|v_{r}| < 1500 \,{\rm kms}^{-1}$
for A2163 (\cite{Holzapfel96b}).
Bahcall \ea (1994) make theoretical
predictions of the one-dimensional
RMS cluster peculiar velocities for a variety of cosmological models.
The results range from $268\,{\rm kms}^{-1}$ for $\Omega =0.3$ cold dark
matter to $614\,{\rm kms}^{-1}$ for $\Omega =1.0$ hot dark matter.
We expect the S-Z results to reach this level of accuracy
soon; until they do, we adopt these theoretical limits to constrain
the peculiar velocity.
For the largest predicted peculiar velocities, we find
$\D y_0/y_0 = \pm 7.4\%$ and $\D H_0/H_0 = ^{+16.6}_{-13.3}\%$.

\subsection{Isobaric Inhomogeneities}
Clumping of the IC gas on sufficiently short scales
could escape detection with the ROSAT/PSPC.
Hydrodynamical simulations of subclustering by Inagaki \ea (1995)
indicate that clumping could result in an overestimation of $H_0$
by as much as $10-20\%$.
Here we consider to what extent the existence of such clumps could
effect the measurement of $H_0$ for A2163.

We assume a simple model for the clumping
consisting of a two-phase medium in pressure equilibrium:
a hot phase of temperature $kT_{h}$, and cooler clumps overdense by a
factor $B$ with respect to the hot phase. The clumps are assumed to be
uniformly distributed in volume with filling factor $f$.
In principle, such clumping can be constrained by X-ray spectral data; 
this is equivalent to placing limits on the superposition of two
isothermal models with the same form for their average density profile
and temperatures $kT_{h}$ and $kT_{c} = kT_{h}/B$.
We define a normalization factor $N_C \propto n^{2}_{e0_h}~kT_{h}^2$,
where $n_{e0_h}$ is the central electron density of the hot phase.
$N_C$ is normalized to the best fit value obtained with the isothermal
model; in the presence of clumping, $H_0/H_{0,iso} = N_C$.

\begin{figure}[htb]
\plotone{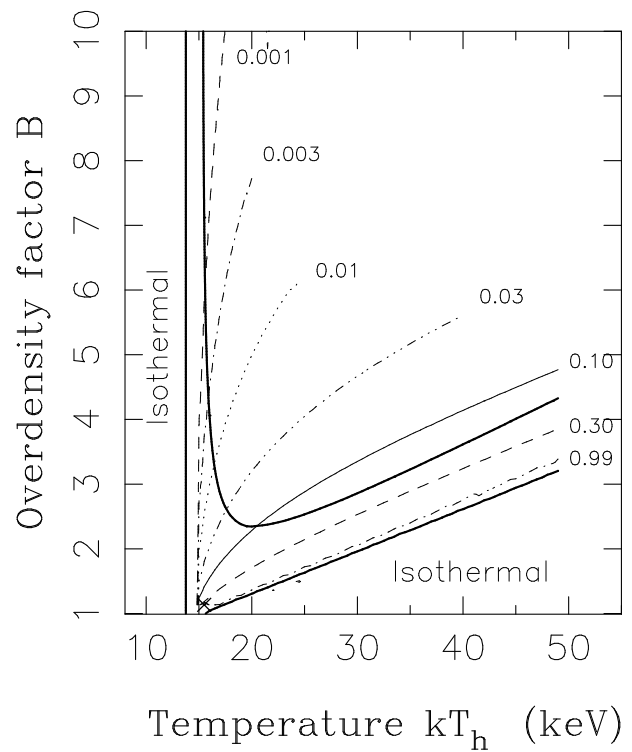}
\caption[]
{Contour map for the two parameters $kT_h$ and $B$ of the
isobaric clumping model.
$kT_h$ is the temperature of the hot phase and $B$ is
the overdensity factor of the clumps.
The heavy line represents the limits at $68\%$ confidence, and is
determined by fitting the GINGA + ROSAT/PSPC spectral data.
Contours of equal filling factor $f$ are indicated by the broken lines.
The best fit point is indicated by a cross.}
\label{kTh_B}
\end{figure}

\begin{figure}[htb]
\plotone{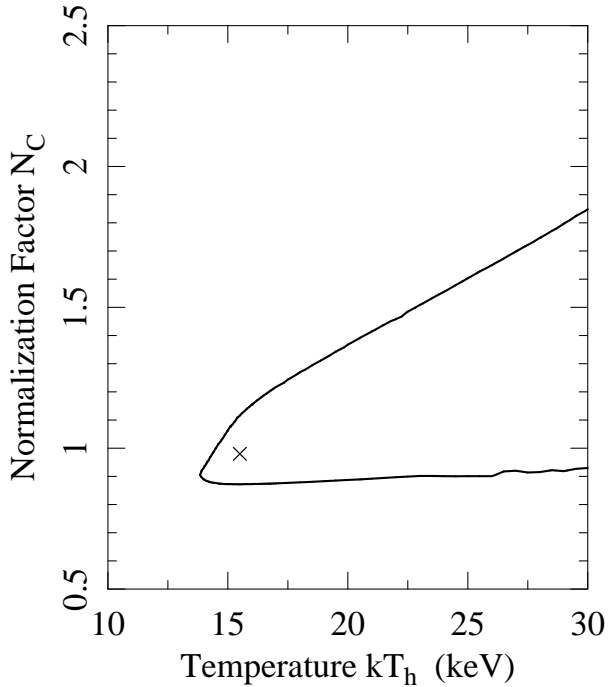}
\caption[]
{Contour map for the two parameters $kT_h$ and $N_C$.
The normalization factor, $N_C$, is defined such as $N_C=H_0/H_{0,iso}$.
A heavy line marks the limits at $68\%$ confidence.
The best fit point is indicated by a cross.}
\label{N_C}
\end{figure}

We fit the GINGA + ROSAT/PSPC spectral data with this model, leaving
$f$, $B$, $kT_{h}$, and $N_C$ free to vary.
The hydrogen column density and the iron
abundance are also left free to vary, but are constrained to have the same
value for the two phases.
The relative normalization between the PSPC and GINGA
data is fixed to the best value obtained from the isothermal
model.
The best fit is obtained for a nearly isothermal model: $kT_{h}
=15.5\,$keV, $B=1.15$, and $N_C=0.98$,  with no significant
improvement in the fit as compared to the isothermal model.
The $68\%$ confidence contours for the two
parameters $kT_{h}$ and $B$ are plotted in Fig.~\ref{kTh_B},
with lines corresponding to equal $f$ values superimposed.
In Fig.~\ref{N_C} we have plot the $68\%$ confidence contours for
the two parameters
$kT_{h}$ and $N_C$; it is apparent that the data prohibit
low values of $N_C$.

In the extreme case of large k$T_h$ values, the gas must contain
clumps at temperature $\sim kT_{iso}$ embedded in a significantly
hotter phase.
In the GINGA energy band, the emissivity
is a weak function of the temperature for $kT \geq 20 \,$keV;
therefore, it is difficult to constrain high values for $kT_h$.
However, large values of $kT_h$ are unlikely.
Unless electron thermal conduction is strongly suppressed by IC
magnetic fields, the clumps would evaporate
on a time-scale much shorter than the Hubble time.
The evaporation of clumps in a hot phase has been studied by 
Cowie \& McKee (1977) and 
Balbus \& McKee (1982).
When the scale of the
thermal structure is less than the electron mean free path,
the evaporation is governed by the saturation parameter, $\sigma_0$.
At a redshift of $z=.2$ and $H_0=50\,{\rm kms}^{-1}{\rm Mpc}^{-1}$ 
one half of the ROSAT/PSPC PSF FHWM is $50\,$kpc.
For clump radii smaller than this scale and density less than
$6 \times 10^{-3}\,{\rm cm}^{-3}$, the conduction is
saturated $(\sigma_0 >1)$.
The evaporation time is then given by
\begin{equation}
t_{evap} = 1.0 \times 10^7\, B \left(\frac{kT_h}{15\,{\rm
keV}}\right)^{-1/2}\left(\frac{R_c}{50 {\rm kpc}} \right) \sigma_0^{-0.209} \, {\rm yr} \, , 
\end{equation}
which is always much shorter than the Hubble time.

Considering only the effect of clumps
colder than the bulk of the IC gas, we determine
$\D H_0/H_{0,iso} = -10.0\%$ at $68\%$ confidence.
We have not treated the case in which the modeled IC gas
also exhibits large scale thermal structure;
we expect the restrictions on the presence of cool clumps
to be even tighter in this case.
It must be pointed out that these limits apply only to IC gas
which is static and free of strong magnetic fields, IC gas that 
does not satisfy these conditions could be considerably more clumped.

\subsection{Astrophysical Confusion}
Although astrophysical confusion from randomly distributed sources is
expected to be small at millimeter wavelengths (\cite{FL}), the
possibility of confusion contributing a considerable systematic error
to the determination of $y_0$ must be considered.
In this region of the sky, the amplitude of anisotropic interstellar dust 
emission calculated from the
IRAS $100\, \mu$m map and scaled with the sky-average spectrum of dust
emission (\cite{Wright91}) is negligible.
A VLA search towards A2163 shows evidence of a radio source $0.8\pr$ west
of the cluster center with an inverted spectrum (\cite{HB92}).
For this source, the flux rises from $1$ to $3\,$mJy between $6$ and
$2\,$cm, suggesting a flux as large as $30\,$mJy at $2.1\,$mm.
However,
Fischer \& Radford report an upper limit of $5\,$mJy
($2\sigma$ in a $20\2pr$ $\delta$, $10\2pr$ RA beam) on point
source emission within
$1\pr$ of the X-ray center at a wavelength of $3.3\,$mm.
For the adopted hybrid model, a $5\,$mJy point source at
$2.1\,$mm would cause the underestimation
of the peak \comp\ by $\D y_0 = +0.24 \times 10^{-4}$.
The sign reflects the fact that the only known radio
source in the scan with a flux and spectrum such that
it could contribute significantly is sufficiently close to the S-Z center
that it would decrease the size of the measured S-Z decrement.
A2163 exhibits the brightest radio halo yet discovered (\cite{HB95}).
From measurements at $1.5$ and $4.9\,$GHz, the integrated flux
from the radio halo is estimated to be less than $1\,$mJy in the
SuZIE $2.1\,$mm band (\cite{HPC}).
This could contribute at most $\D y_0 = \pm .05 \times 10^{-4}$ to the
peak \comp.
We can use these results to place a conservative limit on the 
contribution of radio confusion to the uncertainty in the 
peak \comp, $\D y_0 = ^{+.25}_{-.05} \times 10^{-4}$.

It is also possible for the measurement of the S-Z effect to be confused
by the presence of primary CMB anisotropies.
Haehnelt and Tegmark (1996) estimated the 
confusion limits 
from primary anisotropies to the determination of cluster peculiar velocities.
We use their results to determine the effect of primary anisotropies
on the measurement of the peak \comp.
For $\Omega = 1$ (CDM) models with $\Omega_{baryon} = .01-.1$, cluster
optical depth $\tau=0.015$, and the SuZIE
beam size, they find $|\D v_{pec}| < 300\,{\rm kms}^{-1}$.
From this, we determine that primary anisotropies contribute an uncertainty of
$\D y_0/y_0 = \pm 3.6\%$ to the peak \comp\ parameter in A2163.

\section{Conclusion}
\label{Conclusion}
We have confirmed the previous detection of the S-Z effect
in A2163 at $2.1\,$mm (\cite{W94}). The 1993 and 1994 data sets have been
analyzed using a relativistically correct treatment for the S-Z effect
and the ROSAT derived density profile.
Assuming an isothermal IC gas at the estimated GINGA+ASCA 
isothermal temperature, $T_e=12.4^{+2.8}_{-1.9}\,$keV, 
and including uncertainties in the 
fit amplitude, baseline, calibration, position, and density model, 
we find $y_0 = 3.73^{+.48}_{-.61}\times 10^{-4}$.
Combining the S-Z and X-ray data 
and including uncertainties due to the S-Z and X-ray
normalizations and isothermal IC gas temperature, we find
$H_0(q_0=\frac{1}{2})=60^{+41}_{-23}\,{\rm kms}^{-1}{\rm Mpc}^{-1}$.

There are indications that the IC gas in A2163 is not isothermal.
Recent ASCA results suggest a dramatic decrease in the temperature of 
the IC gas with radius (\cite{Markevitch96}).
The model used by Hughes \ea (1988) to fit the thermal
structure in the Coma cluster is generalized and used as a two
parameter fit to the temperature profile in A2163.
We use the combined analysis of the S-Z and X-ray surface 
brightnesses to place limits on models for the thermal structure
of the gas.
This is the first application of, what is potentially, a powerful 
probe of the state of the IC gas.
So far, the S-Z measurements lack sufficient sensitivity
to distinguish between the thermal structure indicated by the ASCA 
results and an isothermal IC gas. 

Adopting the ASCA thermal structure, a joint analysis of the
GINGA and ASCA/SIS+GIS thermal structure 
yields a central temperature, $T_{e0} = 13.3^{+2.8}_{-1.7}\,$keV.
Using this thermal structure,
and including uncertainties due to the baseline, 
calibration, and density and temperature models,
we determine $y_0 = 3.07^{+.40}_{-.54} \times 10^{-4}$.
Combining the X-ray and S-Z results
and including uncertainties due to the S-Z and X-ray
normalizations and central temperature, we find
$H_0(q_0=\frac{1}{2})=78^{+54}_{-28}\,{\rm kms}^{-1}{\rm Mpc}^{-1}$.

There are several additional contributions to the uncertainties
in $y_0$ and $H_0$.
It is possible that the cluster gas distribution is aspherical.
We use the X-ray morphologies of a sample of clusters
to estimate this contribution to the uncertainty in $H_0$ determined from
a single cluster.
If the cluster exhibits a significant peculiar velocity there is an
additional kinematic component to the S-Z effect.
We use theoretical estimates of cluster peculiar velocities to constrain
the contribution of the kinematic effect to the uncertainty in
$y_0$ and $H_0$.
An IC gas that contains clumps could bias the value of $H_0$.
The ROSAT/PSPC + GINGA spectrum is used
to constrain the contribution that clumped IC gas could make
to the value of $H_0$.
Finally, it is possible that the S-Z effect is astrophysically confused.
We use theoretical and observational results to limit
the uncertainty associated with astrophysical confusion,
including primordial anisotropies of the CMB.
The contributions to the uncertainty in $y_0$ and $H_0$ due to each
of these sources are listed in Tables~\ref{yut} and \ref{hut}.
Including these additional sources of uncertainty in the analysis
which adopts the ASCA thermal structure, we determine
$y_0 = 3.07^{+.54}_{-.60}\times 10^{-4}$ and
$H_0(q_0=\frac{1}{2})=78^{+59}_{-40}\,{\rm kms}^{-1}{\rm Mpc}^{-1}$.

There is an important distinction to be made between the uncertainties
listed in Table~\ref{hut}, those that are essentially statistical and those
that have a significant systematic component.
Because we have no information about the extent of the IC gas along the
line of sight, the IC gas is assumed to be spherically symmetric.
We must average over a sample of clusters to reduce the
contribution to the uncertainty due to asphericity of the IC gas.
The uncertainties due to cluster peculiar velocities and
primary CMB anisotropies are purely statistical and smaller than
that due to cluster asphericity.
These uncertainties, like the statistical uncertainties in the 
measured data, will be reduced for $H_0$ determined from a sample 
of clusters.

A more serious obstacle to the determination of $H_0$ is the
presense of   
systematic errors which can contribute to the results for each 
of the members of a sample of clusters.
If the IC gas has the thermal structure indicated by the ASCA results
for A2163, assuming the gas to 
be isothermal results in the underestimation of $H_0$ by $\sim 30\%$.
To eliminate this source of uncertainty, the thermal structure of 
each cluster in the sample will have to be determined.
All IC gas will be clumped to some degree; this biases the 
results for $H_0$ to higher values.
The presence of clumps could be better constrained by high resolution
observations of the X-ray surface brightness, and
improved X-ray spectra.
Astropysical confusion of the S-Z effect by compact cluster radio sources 
can be virtually
eliminated by observations with mm-wavelength interferometers.
Uncertainty in the mm-wavelength flux of the planetary calibrators
contibutes $\sim \pm 10\%$ uncertainty to the determination of $H_0$.
The COBRAS/SAMBA satellite 
mission may be capable of relating planetary (and some S-Z) 
fluxes directly to the dipole 
anisotropy, which is known to a precision of $\pm.2\%$ (\cite{Fixsen}).

To determine an accurate value of $H_0$
using the S-Z effect, we must reduce the potential sources of 
systematic error to below the level of the statistical uncertainties 
when averaged over a reasonable sample of clusters.
Spatially resolved X-ray spectra obtained by the next generation 
of X-ray satellites will be essential to achieve this goal. 
Using the S-Z effect, it should then be possible to determine $H_0$ 
to an accuracy of $\lesssim 10\%$ with a sample of $\sim 25$  clusters.

We would like to thank Maxim Markevitch and Koujun Yamashita for 
informative discussions concerning the ASCA results.
The referee, Mark Birkinshaw, contributed several helpful comments which
greatly improved the presentation of the results.
Thanks to Antony Schinkel and the entire staff of the CSO for their 
excellent support during the observations.
The CSO is operated by the California Institute of Technology under
funding from the National Science Foundation, contract \#AST-93-13929.
This work has been made possible by a grant from the David and Lucille
Packard foundation and by National Science Foundation grant 
\#AST-95-03226.

\appendix
\section{Appendix: Residual Common-mode Signal}
\label{RSC}
In this appendix, we discuss the removal of the residual common-mode signal which
is not removed in hardware due to the mismatch in the responsivities of the 
detectors in a difference.
A mismatch in responsivities of $\sim 1\%$ can contribute
significant low frequency noise to a detector difference.
Changes in the background loading of the detectors over the course
of an evening can lead to responsivity mismatches of this size.
Removing the residual common-mode signal typically reduces the
noise determined from the distribution of scan amplitudes by
$\sim 10-20\%$ with no significant effect on the fit amplitudes.

For each difference, we determine the average common-mode 
correlation coefficients (${\bar \alpha_{k}}$) by averaging 
$\alpha_{kj}$ over the scans in a given
period of observation; they are typically of order unity.
Because of the difference in electronic gains between
the single and differential signals, the
amplitude of the source in the average single detector signal is 
smaller than in the differential signal by a factor
$G_s/G_d = 1/120$.
If $s_{ji}$ and $d_{kji}$ were completely correlated
we would remove ${\bar \alpha}(G_s/G_d)\sim 1\%$ of the
differential signal.

In a drift scan across a stationary source, the difference
of the signals from two perfectly matched detectors is
orthogonal to their sum.
The same is true for the signals
($d_{31}$, $t_{123}$, $d_{64}$, and $t_{456}$) when
compared to the average of all six single detector signals.
The fraction of average single detector signal to which the differences
are sensitive is proportional to the fractional mismatch
of the detector differences, which is $\sim {\bar \alpha}(G_s/G_d)$.
We estimate the amount of signal removed from the
differential signals as the product of the size of the
subtracted signal and its coupling to the detector
differences to be
$\D y/y \approx \left[{\bar \alpha}\, (G_s/Gd)\right]^2 \sim 10^{-4}$.
Subtraction of the residual common-mode signal has a negligible 
effect on the determination of the source amplitudes.


\begin{thebibliography}{}
\bibitem[Alsop \ea 1992]{Alsop} Alsop, D. C., Inman, C., Lange, A. E., \& Wilbanks, T. 1992, Applied Optics, 31, 6610
 
\bibitem[Arnaud \ea 1992]{Arnaud92} Arnaud, M., Hughes, J. P., Forman, W., \& Jones, C. 1992, \apj, 390, 345
 
\bibitem[Avni 1976]{Avni} Avni, Y. 1976, ApJ, 210, 642
 
\bibitem[Balbus \ea 1982]{Balbus} Balbus, S. \& McKee, C. 1982, \apj, 252, 529
 
\bibitem[Bahcall \ea 1994]{Bahcall94} Bahcall, N. A., Gramman \& M., Cen, R. 1994, \apj, 436,
23
 
\bibitem[Birkinshaw \& Hughes 1994]{BH94} Birkinshaw, M. \& Hughes, J. P. 1994, \apj, 420, 33
 
\bibitem[Birkinshaw, Hughes \& Arnaud 1991]{BHA} Birkinshaw, M., Hughes, J. P., \& Arnaud, K. A. 1991, \apj, 379, 466
 
\bibitem[Cavaliere \ea 1979]{Cal79} Cavaliere, A., Danese, L. \& DeZotti, G. 1979, \aap, 75, 322
 
\bibitem[Cavaliere \ea 1976]{Cal76} Cavaliere, A. \& Fusco-Femiano, R. 1976, \aap, 49 137
 
\bibitem[Chandrasekhar 1950]{Chand} Chandrasekhar, S. 1950 Radiative Transfer, p. 17 (New York: Dover)
 
\bibitem[Church \ea 1996]{Church} Church, S. E., Ganga, K. M., Holzapfel, W. L., Ade, P. A. R., Mauskopf, P. D., Wilbanks, T. M. \& Lange A. E. 1997, \apj, in press 
 
\bibitem[Cowie \& McKee 1977]{Cowie} Cowie, L. \& McKee, C. 1977, \apj, 211, 135
 
\bibitem[Elbaz \ea 1995]{Elbaz} Elbaz, D., Arnaud, M., \& Bohringer, H. 1995, \aap, 293, 337
 
\bibitem[Fischer \& Lange 1993]{FL} Fischer, M. L. \& Lange, A. E. 1993, \apj, 419, 194
 
\bibitem[Fischer \& Radford 1993]{FR} Fischer, M. L. \& Radford, S. J. E. 1993, private communication
 
\bibitem[Fixsen \ea 1996]{Fixsen} Fixsen, D. J. \ea 1996, preprint
 
\bibitem[Glezer \ea 1992]{Glezer} Glezer, E. N., Lange, A. E., \& Wilbanks, T. M. 1992, Applied Optics, 31, 7214
 
\bibitem[Griffin \& Orton 1993]{Griffin} Griffin, M. J., \& Orton, G. S. 1993, Icarus, 105, 537
 
\bibitem[Gunn 1978]{Gunn} Gunn, J. E. 1978, in Observational Cosmology, ed. A Maeder, L. Martinet \&  G. Tammann (Sauverny: Geneva Observatory), 1
 
\bibitem[Haehnelt \& Tegmark 1996]{Haehnelt96} Haehnelt, M. G. \& Tegmark, M. 1996, \mnras, 279, 545
 
\bibitem[Herbig \& Birkinshaw 1992]{HB92} Herbig, T. \& Birkinshaw, M. 1992, private communication
 
\bibitem[Herbig \ea 1995]{Herbig} Herbig, T., Lawrence, C. R., Readhead, A. C. S., \& Gulkis, S. 1995, \apj, 449, L5
 
\bibitem[Herbig \& Birkinshaw 1995]{HB95} Herbig, T. \& Birkinshaw, M. 1995, \baas, 26, 1403
 
\bibitem[Herbig 1995]{HPC} Herbig, T., private communication 1995
 
\bibitem[Holzapfel \ea 1996a]{Holzapfel96a} Holzapfel, W. L., Wilbanks, T. M., Ade, P. A. R., Church, S. E., Fischer, M. L., Mauskopf, P. D., Osgood, D. E., \& Lange, A. E. 1997, \apj, in press
 
\bibitem[Holzapfel \ea 1996b]{Holzapfel96b} Holzapfel, W. L., Ade, P. A. R., Church, S. E., Mauskopf, P. D., Rephaeli, Y., Wilbanks, T. M. \& Lange, A. E. 1997, \apj, in press 
 
\bibitem[Hughes \ea 1988]{Hughes88} Hughes, J. P., Gorenstein, P., \& Fabricant, D. 1988, \apj, 329, 82
 
\bibitem[Hughes \ea 1993]{Hughes93} Hughes, J. P., Butcher, J. A., Stewart, G. C., \& Tanaka, Y. 1993, \apj, 404, 611
 
\bibitem[Inagaki \ea 1995]{Inagaki} Inagaki, Y., Suginohara, T., \& Suto, Y. 1995, PASJ, 47, 411
 
\bibitem[Jones \ea 1993]{Jones} Jones, M. \ea 1993, Nature, 365, 320
 
\bibitem[Kompaneets 1957]{Komp} Kompaneets, A. S. 1957, Soviet Physics-JETP, 4, 730.
 
\bibitem[Markevitch \ea 1996]{Markevitch96} Markevitch, M., Mushotzky, R., Inoue, H., Yamashita, K., Furuzawa, A., \& Tawara, Y. 1996, \apj, 456, 437
 
\bibitem[Mather \ea 1994]{Mather} Mather, J. C. \ea 1994, \apj, 420, 439
 
\bibitem[McMillian \ea 1989]{McMillian} McMillian, S. L. W., Kowalski, M. P., \& Ulmer, M. P. 1989, \apjs, 70, 723
 
\bibitem[Myers \ea 1995]{Myers} Myers, S. T., Baker, J. E., Readhead, A. C. S., \& Herbig, T. 1995, preprint
 
\bibitem[Rephaeli \& Lahav 1991]{RL} Rephaeli, Y. \& Lahav O. 1991, \apj, 372, 21
 
\bibitem[Rephaeli 1995a]{R95a} Rephaeli, Y. 1995a, \araa, 33, 541
 
\bibitem[Rephaeli 1995b]{R95b} Rephaeli, Y. 1995b, \apj, 445, 33
 
\bibitem[Soucail \ea 1995]{Soucail} Soucail G., Arnaud M., \& Mathez G. 1996, in preparation

\bibitem[Squires \ea 1997]{Squires} Squires, G., Neumann, D. M., Kaiser, N., Arnaud, M.,
Babul, A., Bohringer, H., Fahlman, G., \& Woods, D. 1997, \apj, submitted
 
\bibitem[Sunyaev \& Zel'dovich 1972]{SZ72} Sunyaev, R. A. \& Zel'dovich, Ya. B. 1972, Comments on Astrophysics and Space Physics, 4, 173
 
\bibitem[Sunyaev \& Zel'dovich 1980]{SZ80} Sunyaev, R. A. \& Zel'dovich, Ya. B. 1980, \mnras, 190, 413.
 
\bibitem[Wilbanks \ea 1994]{W94} Wilbanks, T.  M., Ade, P. A. R., Fischer, M. L., Holzapfel, W. L., \& Lange, A. E. 1994, \apjl, 427, L75
 
\bibitem[Wright 1979]{Wright79} Wright, E.L. 1979, \apj, 232, 348
 
\bibitem[Wright \ea 1991]{Wright91} Wright E.L. \ea 1991, \apj, 381, 200
 
\bibitem[Zel'dovich \& Sunyaev 1968]{ZS68} Zel'dovich, Ya. B. \& Sunyaev, R. A. 1968, \apss,
4, 301
\end{thebibliography}
\end{document}